# From Routing Delay Shifts to Silent Data Corruption: Neutron-Induced SEU Effects in AXI-Based Zynq UltraScale+ MPSoCs

Mostafa Darvishi, *Senior Member, IEEE*

***Abstract*— SRAM-based FPGA system-on-chip devices are vulnerable to single-event upsets (SEUs) in configuration memory, which may perturb programmable routing resources and degrade communication fabrics. In modern Zynq UltraScale+ MPSoCs, such routing disturbances can introduce small propagation delay shifts that remain logically transparent yet compromise AXI-based data transfers and lead to silent data corruption. Although routing delay degradation and AXI interconnect failures have been studied independently, their experimental correlation under neutron irradiation has not been established. This work presents a cross-layer investigation on a ZCU104 platform integrating routing-dominated delay sensors with an AXI interconnect benchmark comprising replicated accelerators. Neutron irradiation experiments were conducted on the fully operational system, while a frame-level configuration fault injector implemented via the internal configuration access port enables controlled upset emulation. Measured routing delay events are statistically correlated with communication failures, and cross-sections for both timing shifts and AXI malfunctions are derived. The results experimentally demonstrate how neutron-induced routing perturbations propagate into system-level silent data corruption in UltraScale+ MPSoCs, providing insight for resilience-oriented AXI-based design in neutron-rich environments.**

## I. INTRODUCTION

SRAM-based field-programmable gate arrays (FPGAs) have become fundamental components in aerospace, high-energy physics, avionics, and other radiation-prone domains due to their architectural flexibility and capacity for high-performance heterogeneous integration [1]. The emergence of system-on-chip (SoC) devices such as the Zynq UltraScale+ MPSoC has further expanded this applicability by integrating programmable logic (PL) with multi-core processing subsystems and hardened peripherals within a single device [2]. This architectural convergence enables tightly coupled accelerator-based computing platforms but simultaneously introduces complex interdependencies between programmable routing, communication fabrics, and computation blocks. In neutron-rich environments, the configuration memory that defines PL and routing resources remains susceptible to single-event upsets (SEUs), which can alter device behavior in ways that are both subtle and system-critical [1], [3]-[5].

In SRAM-based FPGAs, routing resources are controlled by configuration bits that enable or disable programmable interconnection points (PIPs) [1], [4]. A neutron-induced/fault-injected upset affecting one of these bits may activate an unintended connection or deactivate an intended one. When an unintended PIP becomes active, additional parasitic capacitance and resistance can be introduced into a signal path [4], [5]. Even if logical connectivity remains functionally correct, these parasitic effects may alter signal propagation delay. Recent irradiation studies have demonstrated that such routing perturbations can produce measurable picosecond-scale delay shifts when observed through carefully instrumented ring oscillators (ROs) or timing-sensitive paths. These delay variations constitute a physical manifestation of configuration-level disturbances and represent a mechanism by which radiation effects can influence circuit timing without immediately producing a logical fault [1], [3], [5], [6]

At the architectural level, modern FPGA-based SoCs frequently employ the Advanced eXtensible Interface (AXI) as the primary communication fabric between processing systems and PL accelerators [7]. The AXI interconnect arbitrates transactions, manages data flow, and ensures protocol compliance across multiple masters and slaves [8]. In performance-oriented designs, accelerators are often replicated to increase throughput or provide redundancy. However, these replicated accelerators typically share a common AXI interconnect infrastructure, making the communication fabric a potential single point of failure. Configuration perturbations within the interconnect routing region can therefore propagate beyond localized timing changes and manifest as transaction timeouts, corrupted data transfers, or silent data corruption (SDC) in which incorrect results are delivered without triggering explicit fault flags [1], [4], [8].

Although routing delay shifts [1], [3], [5] and AXI interconnect failures [7], [8] have each been investigated in isolation, their causal relationship under neutron irradiation has not been experimentally established. Fault injection studies have demonstrated that configuration upsets can disrupt AXI communication, but these studies typically rely on emulated bit-flips rather than particle-beam exposure and do not incorporate concurrent timing observability [7]-[9].

Mostafa Darvishi is with Electrical Engineering Department of École de technologie supérieure (ÉTS), Montreal, Canada. He is also VP of Engineering at Evolution Optiks R&D Inc. (e-mail: darvishi@ieee.org).

Conversely, irradiation-based timing studies have characterized routing delay degradation but have not connected these physical effects to higher-level communication failures in a fully operational SoC architecture [1], [3]-[5]. Consequently, it remains unclear whether measurable routing delay shifts can serve as precursors, correlates, or predictors of SDC and availability failures in AXI-based systems implemented on modern UltraScale+ devices [10].

This work addresses this unresolved question through a cross-layer experimental methodology applied to a Zynq UltraScale+ MPSoC implemented on the ZCU104 development platform. The implemented design integrates routing-dominated delay sensors colocated with an AXI interconnect benchmark containing replicated accelerators, thereby enabling simultaneous monitoring of timing perturbations and communication-level outcomes [10], [11]. A frame-level configuration fault injector is incorporated within the design using the Internal Configuration Access Port (ICAP) in conjunction with the Xilinx Soft Error Mitigation (SEM) infrastructure, allowing controlled and reproducible emulation of configuration upsets in addition to neutron-induced events [9], [12]. Neutron irradiation experiments were conducted on the fully functional system, ensuring that both timing and communication behaviors are evaluated under realistic particle exposure conditions [2], [10].

By correlating measured routing delay events with observed AXI transaction failures, this study establishes an experimentally validated connection between configuration-level routing disturbances and system-level communication degradation. Cross-sections for routing delay shifts and communication failure categories are derived, and the temporal proximity between timing perturbations and SDC events is analyzed. Furthermore, the methodology provides a framework for assessing mitigation strategies aimed at reducing the propagation of routing-induced disturbances into communication failures [9], [13].

Through this integrated experimental investigation, the work contributes a physically grounded understanding of how neutron-induced SEUs in programmable routing fabrics can evolve from localized delay shifts into SDC within AXI-based UltraScale+ MPSoCs. This insight is essential for the design of resilient FPGA SoCs intended for deployment in neutron-rich operational environments.

## II. BACKGROUND AND RELATED WORKS

Radiation effects in SRAM-based FPGAs have traditionally been discussed in terms of functional disruption caused by configuration-memory corruption, because the configuration state defines both the implemented logic and the routing connectivity [1], [5], [14], [15]. What has become increasingly clear in modern devices is that radiation-induced configuration upsets can manifest along two tightly coupled but often separately studied dimensions: *first*, as physical perturbations of programmable routing resources that measurably alter timing behavior [1], [3]-[5], and *second*, as architectural-level failures that compromise system operation, sometimes without producing obvious halting symptoms [13], [16].

A key contribution to the timing-focused view was provided in [1], [3]-[5], which experimentally demonstrates that ionizing radiation, including neutron-induced secondary charged-particle effects can induce measurable delay changes in SRAM-FPGA routing networks through configuration-bit upsets that affect PIPs and related routing resources. These works show that routing perturbations can be observable at very fine resolution using appropriately designed sensor structures [3], and it motivates the interpretation of routing delay shifts as physically meaningful events rather than merely indirect artifacts [5]. However, those studies remain principally focused on the routing network itself and do not attempt to connect these delay perturbations to the behavior of complex system communication infrastructures or to quantify the impact on transaction-level correctness in a functional SoC workload.

A complementary system-level perspective emerges from the two AXI Interconnect studies in [7], [8]. In [7], the authors present a reliability evaluation of the AXI Interconnect IP core within Zynq-class AP-SoCs using controlled configuration fault injection and a benchmark that exercises accelerator communication through AXI. The work establishes that the interconnect infrastructure can behave as a critical point of failure, particularly in architectures that attempt to harden computation blocks while leaving the shared communication backbone unhardened. This observation is important because it highlights a structural vulnerability that is not resolved by replicating accelerators alone when the interconnect remains common and can induce correlated failure behavior across replicated datapaths.

In [8], the same research line is extended to a deeper characterization of the failure manifestations observed at the data level, with particular emphasis on the nature of erroneous transferred words under injected configuration disturbances. The work reports that a substantial fraction of observed data errors exhibits stuck-at-like behavior affecting repeated transfers and can appear concurrently across communications with multiple hardware cores connected through the same interconnect. This result is especially consequential for resilience because it indicates a pathway to SDC in which transactions continue to complete while the delivered values are incorrect, thereby evading detection mechanisms that rely on explicit stalls, timeouts, or protocol-level error responses. Together, [7] and [8] establish that AXI interconnect integrity is a first-order concern for accelerator-based systems and that the observable outcomes include both availability failures and silent corruption modes.

While [7] and [8] provide strong evidence that configuration disturbances can provoke severe AXI-level failures, their methodology is primarily based on configuration fault injection rather than particle-beam exposure, and it does not incorporate direct observability of physical timing degradation in the interconnect routing fabric [17]. This distinction matters because the operational environment addressed in this paper is neutron irradiation, where the upset process is stochastic and may produce a spectrum of perturbations, including those that change parasitic loading and delay without necessarily inducing immediate functional disconnection. Without timing observability, it is difficult to determine whether silent corruption modes are more strongly associated with delay-margin erosion, with electrical contention from unintended connections [1], [3]-[5], or with other structural effects. Consequently, the causal pathway from routing-level perturbation to AXI-level SDC remains experimentally unresolved under neutron irradiation.

A mitigation-oriented work was represented in [18], which proposed a domains-based isolation design flow (IDF) [19], [20] intended to reduce the complexity of applying isolation constraints while improving resilience of replicated designs to configuration upsets. The study is significant because it illustrates that placement and routing isolation can reduce the susceptibility of multi-module architectures to injected configuration faults [21] by reducing unwanted coupling and fault propagation between domains. However, like the AXI studies, this mitigation evaluation is grounded in injection-based analysis and does not connect mitigation-induced changes in layout and routing structure to the occurrence rate or distribution of measurable routing delay perturbations under any beam irradiation. In other words, it remains unclear whether isolation primarily reduces the probability of functional failure by preventing severe routing conflicts, or whether it also reduces the incidence of smaller parasitic perturbations that manifest first as timing drift and later as silent corruption.

Taken together, the works presented in [1], [7], [8], [18] establish two important but partially disconnected facts: *first*, the routing fabric of SRAM-based FPGAs can exhibit measurable delay shifts under radiation due to upset-induced routing perturbations [1], [5]. *Second*, AXI interconnect infrastructures can experience availability failures and SDC under configuration disturbances, even in architectures that attempt redundancy at the accelerator level [7], [8]. In addition, placement and routing isolation has demonstrated promise as a mitigation approach for multi-module designs under configuration fault injection [21]-[23]. The missing piece is an experimentally validated, cross-layer link connecting routing delay shifts to AXI-level error manifestations under neutron irradiation on a modern MPSoC platform, together with an evaluation of whether isolation reshapes not only the functional error distribution but also the underlying timing-perturbation statistics. Addressing that gap requires a methodology that simultaneously observes routing-level timing behavior and transaction-level correctness in a fully operational system under neutron exposure, while also enabling controlled configuration upset emulation to support reproducibility and mechanism isolation.

## III. PHYSICAL AND ARCHITECTURAL FOUNDATIONS

Understanding how neutron-induced configuration upsets propagate into system-level communication failures requires a structured examination of the intermediate physical and architectural mechanisms involved [3], [24]. While Section II established the existing separation between routing-level timing studies and AXI-level reliability analyses, this section develops the formal framework that connects these domains. The objective is to model, at successive levels of abstraction, how a configuration-memory upset modifies routing connectivity, how that modification perturbs signal propagation delay, how timing margin erosion can influence synchronous communication fabrics [25], and how these effects can ultimately manifest as availability degradation or SDC. By explicitly articulating this physical-to-architectural propagation chain, the subsequent experimental methodology and statistical analysis are grounded in a well-defined mechanistic foundation rather than empirical correlation alone [13], [24].

### *A. Configuration Upsets and Routing Perturbation Mechanisms in UltraScale+ Devices*

In SRAM-based UltraScale+ FPGAs, the logical and structural configuration of the PL fabric is determined by distributed configuration memory organized into frames. Each frame stores a collection of configuration bits that control lookup tables (LUTs), flip-flops, clocking resources, and PIPs.

Routing resources in UltraScale+ devices consist of hierarchical interconnection networks composed of local switch matrices, horizontal and vertical routing channels, and PIPs that selectively connect wire segments. Each PIP is controlled by configuration bits that determine whether a conductive path is enabled. If an upset affects one of these control bits, an unintended connection may be introduced between routing wires, or an intended connection may be removed [25], [5].

As shown in Fig. 1, the most critical scenario for delay degradation arises when an SEU occurs in the configuration memory, it may toggle the stored state of a memory configuration bit that controls a PIP inside a switch matrix. Once the undesired PIP is activated, while the original routing path remains intact, the signal path experiences an additional electrical branch that was not part of the intended design. Even if the unintended wire segment does not carry switching activity, its capacitive coupling to the active net increases the effective load seen by the driving element. The resulting increase in capacitive loading lengthens the signal rise and fall times and increases propagation delay along the path [4], [5].

The magnitude of the delay perturbation depends on the electrical characteristics of the activated routing segment, including its length, the number of downstream switch matrices it spans, and its proximity to other active nets [1]. Because UltraScale+ routing networks are optimized for density and performance, routing segments may span multiple metal layers and clock regions. Consequently, even a single activated PIP may introduce a delay shift measurable at picosecond resolution. Importantly, such a perturbation may not immediately violate logical correctness if the affected path retains sufficient timing margin. Therefore, routing-level

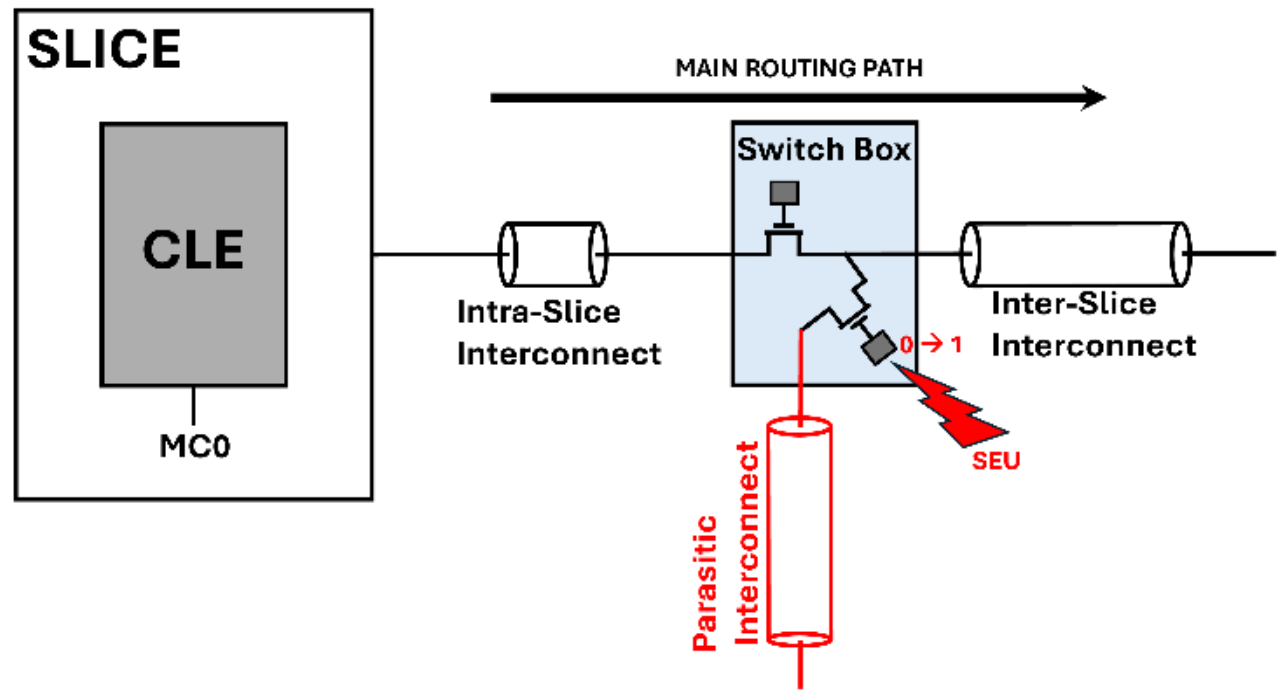

Fig. 1. Conceptual illustration of SEU-induced routing delay degradation in an UltraScale+ FPGA. An induced configuration upset toggles the enable bit of the PIP within a switch matrix, unintentionally connecting a previously unused interconnect segment to the main routing path. The added parasitic loading ($\Delta C$) increases the effective propagation delay ($\Delta T$) while logical connectivity of the primary path remains intact [4].

SEUs can create a class of events characterized by measurable delay change without immediate functional disruption [1], [5], [10], [25]. This mechanism forms the physical basis for considering routing delay shifts as primary radiation-induced events. Rather than treating functional failure as the only relevant manifestation of configuration upset, the routing perturbation itself can be considered a measurable intermediate state that may evolve into higher-level malfunction under certain architectural conditions.

### *B. Timing Margin Erosion and Critical Path Sensitivity*

Digital systems are designed with finite timing margins to tolerate process variation, temperature fluctuation (PVT), and voltage noise [26]-[28]. In synchronous architectures, correct operation requires that data signals arrive at sampling elements within defined setup and hold windows relative to clock edges. When a routing perturbation increases propagation delay along a path that contributes to a critical timing constraint, the available margin between data arrival and the setup boundary is reduced.

If the delay shift remains smaller than the available margin, no immediate violation occurs. However, the erosion of timing margin increases the system's susceptibility to secondary effects such as dynamic voltage variation, clock jitter, or additional radiation-induced perturbations. Should the cumulative delay exceed the setup limit, incorrect sampling may occur. In the context of communication fabrics, such sampling errors can manifest as corrupted data words, malformed control signals, or incorrect sampling of handshake transitions, and in marginal cases metastability at receiving registers [3], [29].

The timing margin erosion process and its relationship to the setup-violation threshold are illustrated in Fig. 2, where the nominal arrival preserves margin while the shifted arrival reduces it and may encroach on the setup window. Under nominal conditions, the data launched by upstream logic arrives at the receiving sampling element sufficiently before the active clock edge ($t_0$), leaving a positive setup timing margin that provides robustness against process variation, environmental changes, and clock uncertainty. When an SEU-induced routing perturbation increases propagation delay by an increment $\Delta T$, the data transition shifts later in time. This shift reduces the distance between the data arrival and the setup boundary defined by the receiving clock edge ($t_1$), thereby eroding the available setup margin. If the delay shift is large enough that the data transition enters the setup window preceding the sampling edge, the receiving register may capture an incorrect value. In communication fabrics, this failure mechanism is particularly important because it does not require a topological disconnection; instead, the interconnect may continue to transfer transactions while silently corrupting data values, which is consistent with SDC rather than a hard availability failure.

It is important to distinguish between deterministic logical disconnection and timing-induced corruption. A disconnection resulting from a configuration upset typically produces a persistent and easily detectable failure, such as a transaction timeout or a permanent stall. In contrast, timing-induced corruption may be intermittent and data-dependent. Because handshake protocols such as AXI decouple address, data, and response channels, subtle timing misalignment may permit transactions to complete while delivering incorrect data values. Therefore, routing delay shifts represent not merely a physical curiosity but a potential precursor mechanism for communication-level silent failures. Establishing this link requires simultaneous observability of routing-level timing behavior and system-level transaction correctness.

### *C. Architectural Vulnerability of AXI Interconnect Fabrics*

AXI protocol organizes communication into multiple independent channels, including separate address, data, and response paths, each governed by valid and ready handshake signaling. Correct AXI operation therefore relies not only on logical connectivity but also on precise temporal coordination among these channels. Within a PL implementation, all AXI signals are conveyed through routing resources defined by configuration memory. The AXI interconnect core performs arbitration, multiplexing, address decoding, and data routing, often spanning substantial routing regions inside the programmable fabric.

In accelerator-based architectures implemented on Zynq UltraScale+ MPSoCs, the processing system typically operates as an AXI *master* while peripherals or hardware accelerators implemented in programmable logic act as AXI *slaves*. When multiple slave IP cores are instantiated, they commonly share a single AXI interconnect backbone. Although the slave modules are logically independent, their communication paths converge within the interconnection routing fabric. This architectural choice introduces a structural concentration of routing resources whose integrity becomes critical for overall system correctness.

The architectural concentration of vulnerability within such a design is illustrated in Fig. 3. The processing system (*AXI Master*) communicates with PL peripherals (*AXI Slave IPs*) through an *AXI interconnect* implemented entirely in the routing fabric. In the illustrated block design, example slave IP cores such as AXI GPIO and AXI UARTLite are connected downstream of the interconnect. While these IP cores perform distinct functions, their AXI channels traverse shared routing regions inside the interconnect.

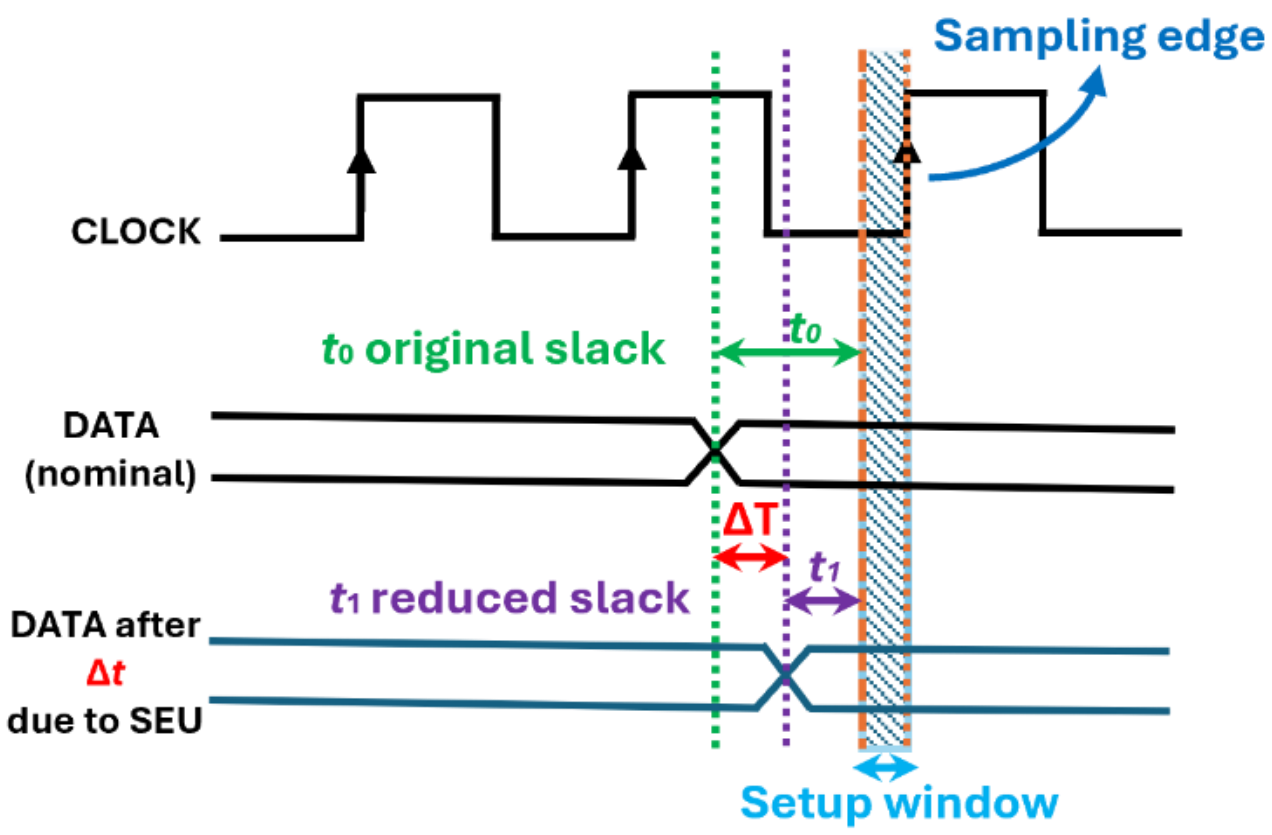

Fig. 2. Timing margin erosion due to an SEU-induced routing delay shift. Under nominal conditions, data arrival at time $t_0$ leaves a positive setup slack relative to the sampling edge. An SEU-induced delay increment $\Delta t$ shifts the arrival to $t_1$, reducing the available slack. If the shifted arrival encroaches upon the setup window preceding the sampling edge, setup timing may be violated, enabling incorrect sampling and potential silent data corruption without requiring a structural interconnect failure [3].

The highlighted region in Fig. 3 represents a shared routing area implemented using configuration-controlled PIPs. An SEU affecting configuration memory bits within this region may perturb routing connectivity or modify signal delay characteristics common to multiple AXI channels. Because the interconnect distributes transactions to all connected slave IPs, a single routing upset in this shared region can influence communication paths to multiple peripherals simultaneously.

The architectural consequences of such an upset depend on its electrical manifestation. If the upset disables a required routing segment, the resulting behavior may be a complete transaction stall, bus error, or system hang, corresponding to availability degradation. However, if the upset activates an unintended routing branch or otherwise alters signal delay without severing connectivity, the impact may be more subtle. Delay shifts affecting data paths or valid/ready handshake signals can disturb temporal alignment among AXI channels while preserving apparent logical connectivity. In this condition, transactions may still complete nominally at the protocol level even though internal timing constraints are violated. Corrupted data values may therefore propagate to the slave IPs without triggering explicit protocol-level error signaling.

Because AXI protocol correctness depends on tightly bounded timing relationships among channels, even small routing-induced delay perturbations can have disproportionate architectural consequences. The AXI interconnect thus provides a structurally sensitive environment in which routing delay degradation can propagate into communication-level malfunction. This architectural sensitivity, combined with the shared nature of the routing fabric, motivates the selection of the AXI interconnect as the focal communication infrastructure in the present study.

### *D. Event Taxonomy and Observable Categories*

To enable rigorous cross-layer analysis, it is necessary to formally define the categories of observable events considered in this study. Because routing-level perturbations and architectural-level malfunctions are not inherently equivalent, a structured taxonomy is required to distinguish physical delay phenomena from communication failures and to enable statistically meaningful correlation analysis.

At the physical layer, a routing delay event is defined as a statistically significant deviation in the measured propagation delay of a monitored routing structure relative to its nominal baseline. Such events reflect configuration-induced perturbations in programmable interconnection resources and may occur independently of immediate functional disruption.

At the architectural level, communication outcomes are classified according to externally observable behavior at the AXI interface. An "**availability failure**" corresponds to a transaction stall, timeout, bus hang, or loss of forward progress in which AXI communication does not complete within defined protocol constraints. In contrast, "**silent data corruption (SDC)**" refers to the completion of an AXI transaction that delivers incorrect data without triggering protocol-level error signals, exceptions, or system resets. "**Detectable protocol errors**" constitute a separate class in which AXI error responses or interrupt mechanisms explicitly signal a malfunction.

Because routing delay shifts and communication failures may not occur simultaneously, correlation analysis requires the definition of a temporal association window. A "**correlated event**" is therefore defined as a communication failure occurring within a predefined time interval following a routing delay event. This definition allows the conditional probability of architectural malfunction given a preceding routing perturbation to be evaluated quantitatively. The formal definitions used throughout this work are summarized in Table I. These categories form the basis for cross-section estimation, statistical confidence analysis, and delay-to-failure correlation modeling in the subsequent sections.

### *E. Statistical Framework and Cross-Section Formulation*

Quantifying neutron-induced effects requires a probabilistic framework that connects observed event counts to particle exposure. In this work, neutron exposure is expressed through the accumulated fluence Φ, defined as the time-integrated neutron flux incident on the device under test, with units of particles per unit area ($n/cm^2$). Observable outcomes are treated as event processes occurring during irradiation, and each event type is counted according to the classification defined in Table I. Each event category is treated as a stochastic process driven by configuration-memory upsets under neutron irradiation.

For an event type $X$, the experimental cross-section $\sigma_X$ is defined as the number of observed events $N_X$ normalized by the accumulated fluence Φ. This definition provides a fluence-normalized measure of susceptibility that is directly comparable across experiments conducted at different flux

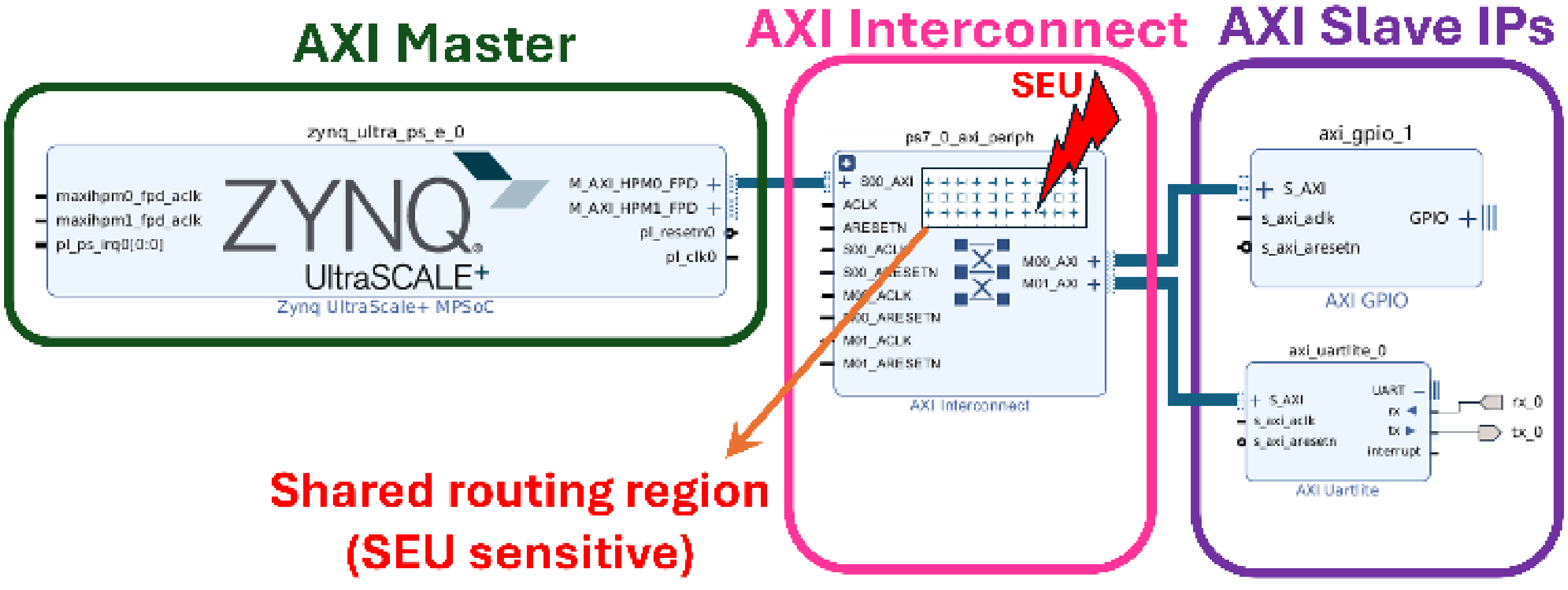

Fig. 3. Architectural vulnerability of a shared AXI interconnect in a Zynq UltraScale+ MPSoC block design. The processing system (AXI master) communicates with PL peripherals through an AXI interconnect implemented in the routing fabric.

Table I. Event classification framework used in this study

| Event Category | Formal Definition | Observable Metric | System-Level Interpretation |
|---|---|---|---|
| **Routing Delay Event** | Measurable deviation of monitored routing propagation delay from nominal baseline due to configuration perturbation | Frequency shift or time-delay variation exceeding defined statistical threshold [1] | Physical-layer manifestation of SEU in routing resources [1], [5] |
| **Availability Failure** | AXI transaction fails to complete within protocol-defined timing constraints or forward progress is lost | Transaction timeout, bus hang, system stall, watchdog trigger | Communication interruption or system-level service degradation |
| **Silent Data Corruption (SDC)** | AXI transaction completes successfully but delivers incorrect data without explicit error signaling | Data mismatch detected by checker logic with no AXI error response | Undetected functional corruption at architectural level |
| **Detectable Protocol Error** | AXI error response or interrupt explicitly indicates communication fault | AXI SLVERR/DECERR response or interrupt assertion | Detected architectural malfunction |
| **Correlated Event** | Communication failure occurring within predefined temporal window following a routing delay event | Temporal coincidence within correlation window $\Delta T$ | Statistical association between routing perturbation and architectural failure |

levels or durations. The cross-section is therefore computed as

$$\sigma_X = \frac{N_X}{\Phi} \quad (1)$$

In the present study, at minimum, two cross-sections are of central interest: the *routing-delay-event* cross-section $\sigma_D$ is obtained using $X = D$, where $N_D$ denotes the number of routing delay events identified by the timing sensor analysis. The *architectural-failure* cross-sections are obtained using $X = F$, where $N_F$ corresponds to a specific communication-level category such as availability failures or SDC. Because these event categories are distinct, each must be computed and reported separately rather than aggregated into a single failure metric.

Since neutron-induced events in configuration memory are rare and stochastic under typical beam flux levels, event counts are modeled as a Poisson process. Under this assumption, the probability of observing $N_X$ events during a fixed fluence is governed by a Poisson distribution whose *mean* is proportional to *fluence*. This model is appropriate for independent upset occurrences and enables principled uncertainty quantification. The uncertainty in the estimated cross-section is therefore captured through confidence intervals on the Poisson count $N_X$, which are then scaled by fluence. In practice, if $[N_X^-, N_X^+]$ denotes a confidence interval for the true Poisson mean count at a chosen confidence level, the corresponding cross-section interval is expressed as

$$\sigma_X^- = \frac{N_X^-}{\Phi}, \qquad \sigma_X^+ = \frac{N_X^+}{\Phi} \quad (2)$$

This formulation ensures that uncertainty estimates remain valid even when event counts are small, which is frequently the case in radiation experiments when focusing on narrowly defined failure categories.

Beyond independent susceptibility characterization, the goal of this work is to evaluate whether *routing delay shifts* are statistically associated with *communication-level malfunctions*. This is addressed using a conditional probability framework. Let $D$ denote the occurrence of a *routing delay event* and $F$ denote the occurrence of an *architectural failure event* of a specified class. A temporal association window of duration $\Delta T$ is defined such that a failure $F$ is considered temporally associated with a delay event $D$ if it occurs within $\Delta T$ following $D$. This definition does not assume causality by itself, but it provides a testable and repeatable criterion for statistical correlation. Using this definition, the conditional probability of failure given a preceding delay event is estimated as

$$\hat{P}(F \mid D) = \frac{N_{F\cap D}}{N_D} \tag{3}$$

where $N_{F\cap D}$ is the number of failures observed within the association window following a delay event, and $N_D$ is the total number of delay events. This estimate becomes meaningful only when $N_D$ is non-zero and sufficiently large to support statistical inference. To avoid misinterpretation, the conditional probability is evaluated in conjunction with the baseline failure probability observed over the experiment's duration. A statistically meaningful association is indicated when the observed failure likelihood following delay events exceeds the baseline level in a manner that cannot reasonably be explained by random coincidence, taking into account uncertainty from finite event counts. The key statistical quantities used throughout this study are summarized in Table II.

Table II. Key Statistical Metrics and Definitions

| Metric | Mathematical Definition | Interpretation in This Work |
|---|---|---|
| Routing delay event cross-section ($\boldsymbol{\sigma_D}$) | $\boldsymbol{\sigma_D} = N_D/\Phi$ | Fluence-normalized susceptibility of routing delay events under neutron irradiation |
| Architectural failure cross-section ($\boldsymbol{\sigma_F}$) | $\boldsymbol{\sigma_F} = N_F/\Phi$ | Fluence-normalized susceptibility of a specified architectural failure class (availability failure or silent data corruption) |
| Conditional failure probability $\boldsymbol{\hat{P}(F \mid D)}$ | $\hat{P}(F \mid D) = N_{F\cap D}/N_D$ | Likelihood of observing an availability failure or silent data corruption within ( \Delta T ) following a routing delay event |

The physical, architectural, and statistical foundations developed in this section establish a structured framework for evaluating how neutron-induced routing perturbations propagate into communication-level malfunction, thereby motivating the experimental implementation and measurement methodology described in the following section.

## IV. IMPLEMENTED SYSTEM AND EXPERIMENTAL SETUP

To experimentally evaluate the propagation chain established in Section III, a dual-board radiation testing platform was implemented using two Xilinx ZCU104 development boards, each featuring a Zynq UltraScale+ MPSoC. One board serves as the irradiated device under test (DUT), while the second board operates as a non-irradiated supervisory and reference system.

### *A. Dual-Board Architecture Overview*

The implemented architecture is shown in Fig. 4. The irradiated DUT contains the AXI communication fabric, ring oscillator sensor array, custom AXI correctness-checking IP, and Soft Error Mitigation (SEM) infrastructure. The second ZCU104 board, located outside the neutron beam, functions as a supervisory controller and reference platform which is not shown here for simplicity. The two boards communicate through a deterministic FMC-based interface, enabling low-latency bidirectional data exchange and synchronized event reporting.

On the DUT, the Zynq UltraScale+ processing system operates as the AXI master and communicates with the PL peripherals through an AXI interconnect instantiated in the fabric. Downstream of the interconnect are multiple AXI slave IP blocks, including AXI GPIO, AXI UARTLite for runtime logging, and a custom AXI IP implementing deterministic correctness verification. A separate AXI UARTLite instance interfaces with the SEM controller for configuration error reporting and controlled injection. The highlighted red shaded region in Fig. 4 denotes the shared routing resources within the AXI interconnect that are susceptible to neutron-induced single-event upsets. These routing structures form the architectural focal point of the study. Clock and reset signals are omitted in Fig. 4 for clarity. In the implemented system, both boards operate in stable clock domains derived from local oscillators, and synchronization between boards is achieved through timestamped event exchange over the FMC interface.

### *B. Functional Roles of the Two Boards*

The irradiated board serves as the DUT and hosts all routing-sensitive structures and AXI communication logic under investigation. It performs continuous AXI transactions under supervision while routing delay is monitored through distributed ring oscillators. The non-irradiated board performs four critical roles: **First**, it generates deterministic traffic stimuli and receives response data from the DUT through the FMC interface. This ensures that expected results are computed externally and are not subject to corruption by the same radiation effects affecting the DUT. **Second**, it acts as a golden reference for SDC detection. A mismatch between the DUT response and the externally computed golden result is classified as SDC when no AXI protocol error is reported. **Third**, it provides continuous supervisory logging. Even if the DUT experiences availability failure or partial system instability, the supervisory board continues recording timestamps, transaction states, and event markers, preventing data loss. **Fourth**, it provides environmental and temporal reference monitoring. Because the supervisory board is not exposed to neutron radiation, its internal oscillators and timing structures serve as reference baselines, enabling discrimination between SEU-induced step-like delay shifts and slow environmental drift. This architectural separation significantly reduces instrumentation-induced ambiguity and strengthens the statistical validity of the correlation framework defined in Section III-E.

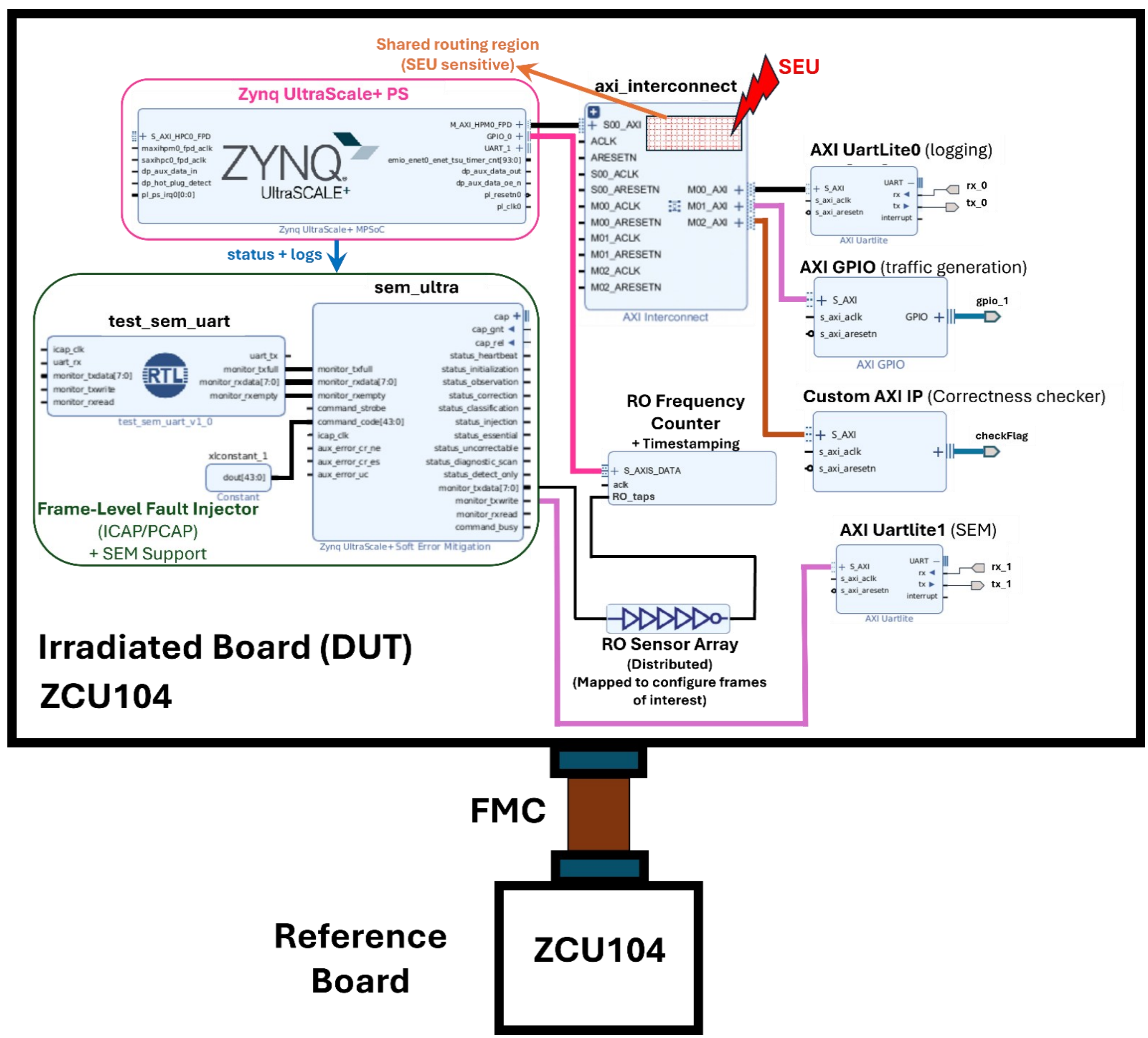

Fig. 4. Implemented architecture on the Xilinx ZCU104 development boards (DUT) exposed to irradiation.

*C. Distributed Routing-Dominated Ring Oscillator Array and Frame Mapping*

To experimentally quantify routing delay perturbations induced by configuration upsets while preserving architectural interpretability, a distributed array of routing-dominated ring oscillators (ROs) was implemented within the PL fabric on the DUT. The objective of this sensing structure is to maximize sensitivity to configuration-controlled interconnects while minimizing the contribution of combinational logic, thereby isolating routing delay events from logic-dependent effects.

The topology of each RO is illustrated in Fig. 5. Each oscillator is composed of two LUTs: one configured as an inverter and the other as an enable-controlled buffer driven by the *EN* signal. The feedback loop is intentionally routed through a chain of switch matrices so that the oscillation period is dominated by programmable routing resources rather than LUT delay. By restricting the oscillator to only two LUTs, the implemented structure minimizes logic depth and ensures that frequency variations are primarily attributable to perturbations in the routing fabric.

Within each switch matrix, the solid blue arrows shown in Fig. 5 represent the intentionally configured routing path of the RO. These segments correspond to the PIPs selected during implementation. In contrast, the dotted lines represent other PIPs associated with the same configuration point but not initially activated in the nominal routing solution. These dormant connections constitute potential parasitic branches. If an SEU flips the corresponding configuration bit, one of these inactive PIPs may become enabled, electrically attaching an unintended branch to the active routing path. Such a perturbation increases parasitic loading and may introduce additional delay without necessarily breaking logical connectivity [1]. The use of different colors for the dotted branches in successive switch matrices improves visual clarity while emphasizing that the same upset mechanism may arise at multiple locations along the routing path.

By forcing the oscillatory loop to traverse multiple switch matrices, the RO becomes strongly sensitive to routing-level perturbations. An upset affecting routing configuration bits therefore modulates the oscillation frequency directly, allowing fine-grained detection of routing delay shifts. The frequency of each RO is measured through dedicated counters and timestamping logic and is compared against its nominal baseline so that statistically significant deviations can be identified as routing delay events.

While Fig. 5 describes the internal topology and upset sensitivity mechanism of a single RO, architectural correlation requires spatial awareness at the configuration-frame level. This spatial relationship is illustrated in Fig. 6, which maps the distributed RO instances to the AXI interconnect routing region. Because the AXI interconnect occupies a concentrated portion of the PL fabric containing dense configuration-controlled routing resources, it represents the most relevant region for studying the propagation of routing perturbations into communication-level failures. Multiple ROs are therefore placed in and around this region using Pblock constraints so that each oscillator is associated with a localized subset of configuration frames.

The left side of Fig. 6 shows the spatial placement of the RO instances within the SEU-sensitive AXI interconnect region, while the right side provides the corresponding mapping between RO identifiers and configuration-frame addresses. This mapping allows each measured frequency deviation to be associated with a specific configuration-frame region, thereby enabling later correlation between routing delay events, injected or radiation-induced configuration upsets, and communication-level outcomes. In addition, the frame allocation table provides the basis for extracting event statistics and cross-sections on a per-frame or per-region basis.

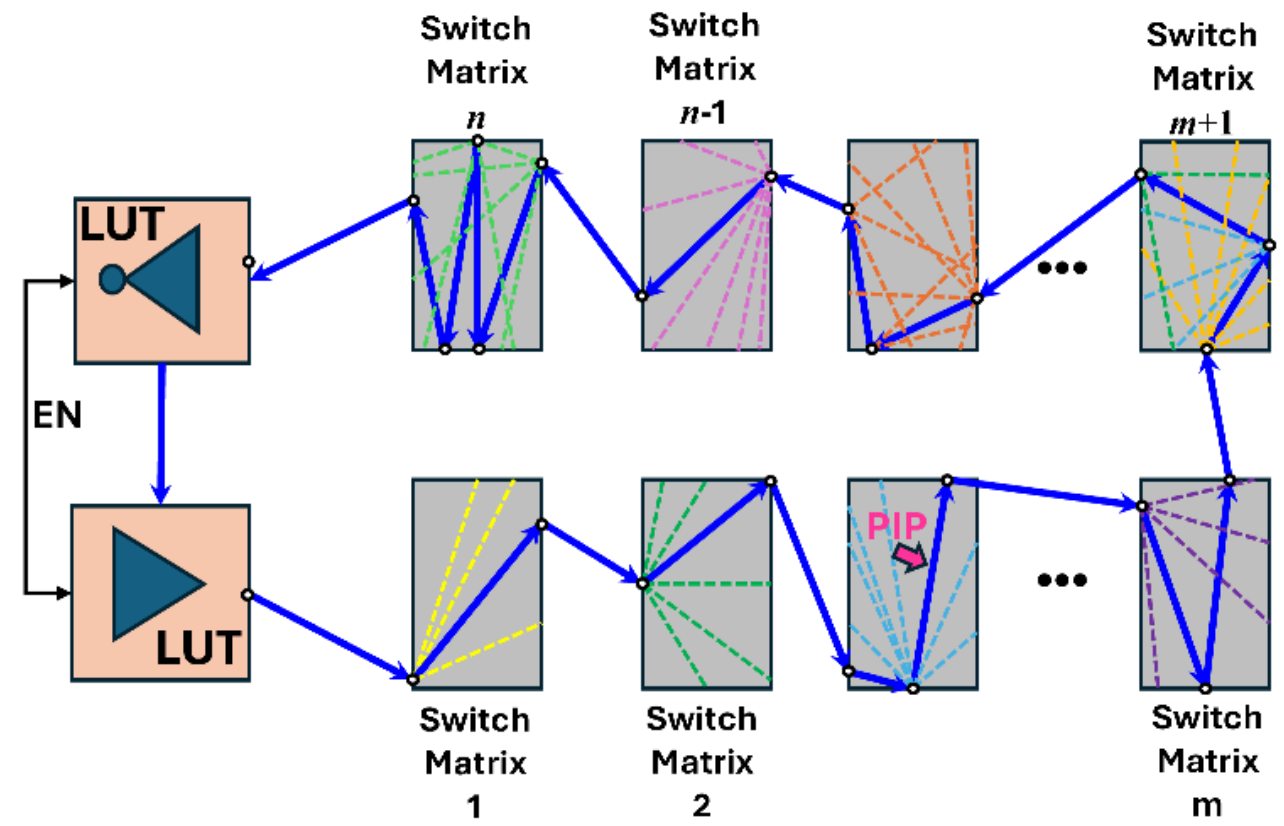

Fig. 5. Routing-dominated ring oscillator topology implemented in the DUT. The oscillator consists of two LUTs (inverter and enable-controlled buffer) and a feedback loop traversing multiple switch matrices. Solid blue arrows denote the configured routing path. Colored dotted lines represent inactive PIPs sharing the same configuration domain that may become parasitic interconnects under SEU. EN enables or disables oscillation.

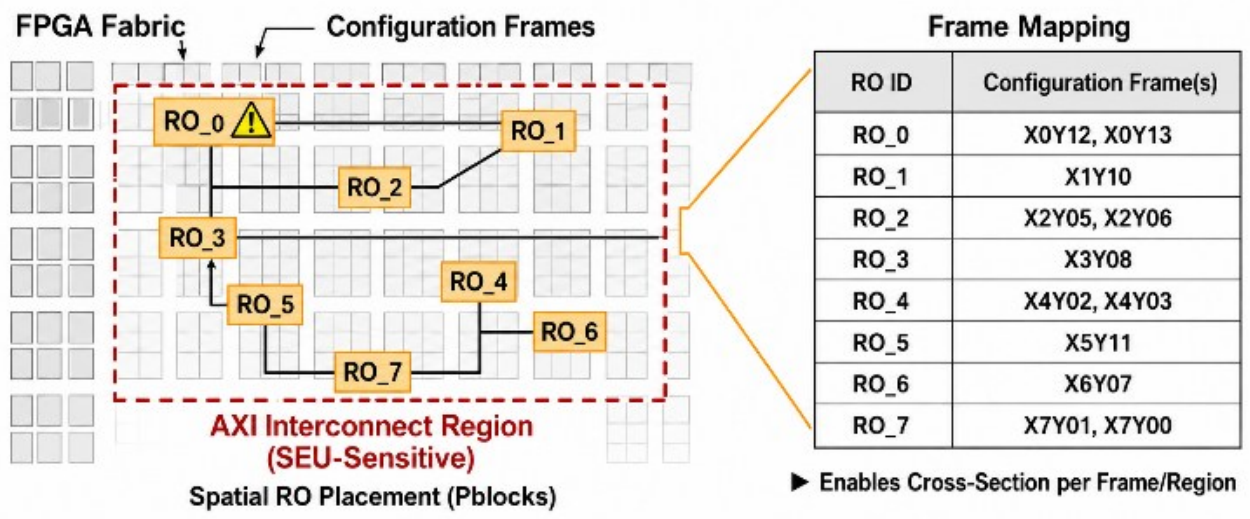

Fig. 6. Spatial mapping of distributed RO instances to configuration-frame regions within the AXI interconnect routing fabric. The left panel shows the placement of the ROs inside the SEU-sensitive interconnect region, while the right panel associates each RO identifier with its corresponding configuration-frame address set. This mapping enables routing-delay events to be correlated with localized frame regions and supports per-frame or per-region event analysis.

### *D. AXI Interconnect Benchmark and Transaction-Level Correctness Monitoring*

After establishing routing-level observability through the distributed RO array and frame mapping described in Subsection IV-C, the experimental platform requires a complementary architectural workload capable of exposing communication-level failure modes. To this end, an AXI-based benchmark was implemented in the PL side of the irradiated DUT. Its purpose is to continuously exercise the shared AXI interconnect under neutron exposure and controlled fault injection while producing deterministic transaction outcomes that can be verified independently. This structure enables simultaneous observation of routing-delay perturbations and communication-level correctness, thereby providing the cross-layer observability required by the objectives of this work. This benchmark concentrates transaction activity within the same routing region monitored by the distributed RO array. As a result, the platform can observe whether routing perturbations detected physically in the vicinity of the interconnect are accompanied by architectural degradation at the transaction level.

The transaction-level monitoring flow is illustrated in Fig. 7. On the irradiated board (DUT), the processing system (PS AXI Master) transmits AXI transactions through the interconnect to a set of replicated AXI slave or accelerator Ips (e.g., Accel $IP_0$, … Accel $IP_N$). The returned data values and AXI response signals are captured by a local checker implemented within the DUT. This checker records transaction completion status, returned data, AXI response behavior, timeout conditions, and local diagnostic flags. These observations are assembled into "*timestamped event records*" and transferred through the FMC link to the non-irradiated supervisory ZCU104 board.

The supervisory board performs independent correctness verification using an external golden reference. This separation is essential because it prevents the correctness reference from being corrupted by the same neutron-induced configuration upsets that may affect the irradiated DUT. For each observed transaction, the supervisory board compares the returned result and communication status against the expected outcome and classifies the event according to the taxonomy defined in Subsection III-D. A transaction is considered *correct* only when it completes within the allowed interval, returns the expected data value, and does not produce an explicit protocol-level error indication.

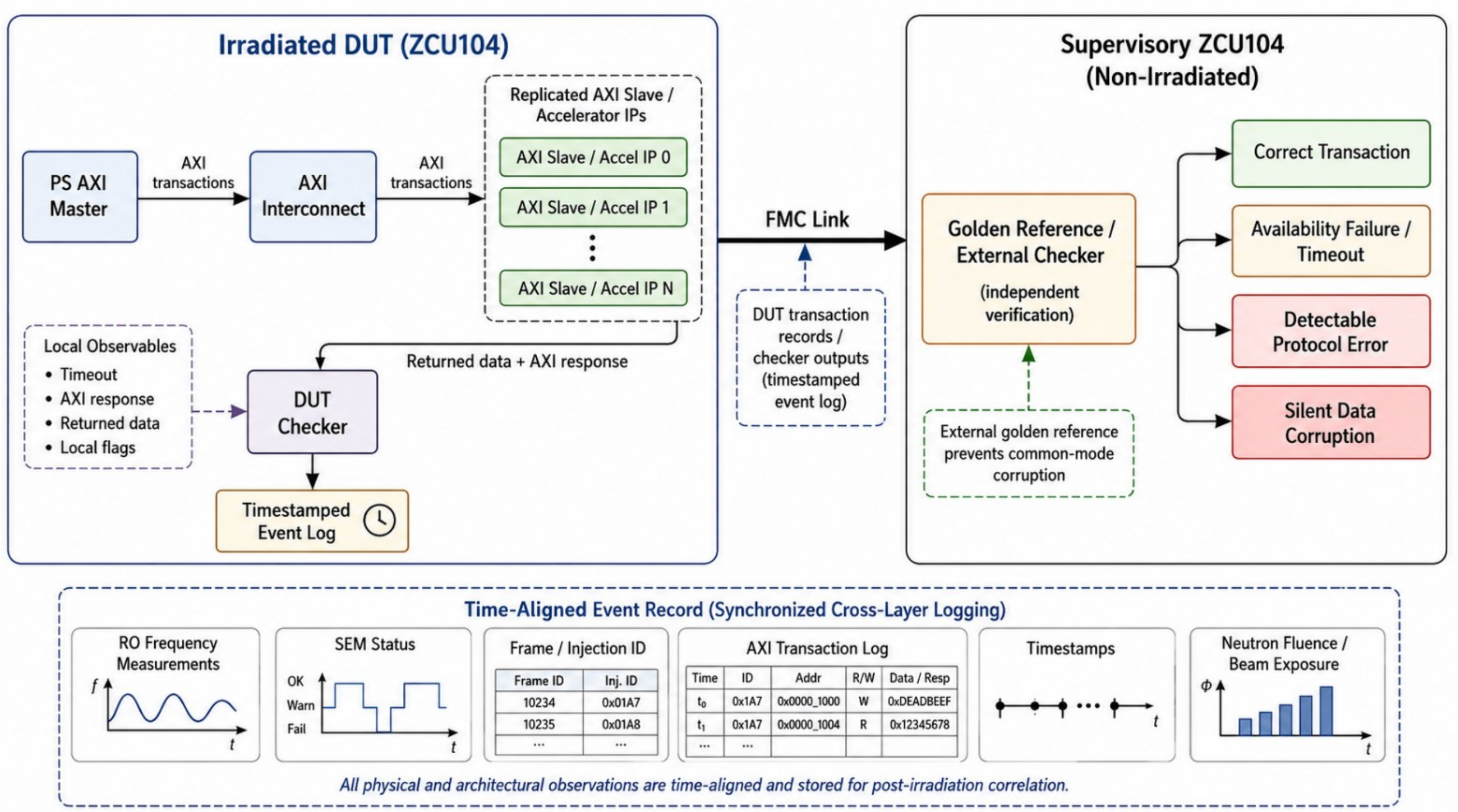

Fig. 7. AXI transaction-level correctness monitoring and synchronized cross-layer event logging. The irradiated DUT issues deterministic AXI transactions from the processing-system AXI master through the shared AXI interconnect to replicated AXI slave or accelerator IPs. Returned data, AXI response signals, timeout status, and local flags are captured by the DUT checker and transferred through the FMC link to the non-irradiated supervisory ZCU104 board. The supervisory board performs independent golden-reference verification and classifies each outcome as a correct transaction, availability failure, detectable protocol error, or silent data corruption. RO frequency measurements, SEM status, frame or injection identifiers, AXI transaction logs, timestamps, and neutron fluence records are stored in a time-aligned event record for post-irradiation correlation.

An *availability failure* is recorded when a transaction fails to complete within the predefined timeout interval, when the interconnect loses forward progress, or when a persistent stall condition is observed. A *detectable protocol error* is recorded when the communication completes abnormally with an explicit AXI error response or another directly observable protocol-level fault indication. A *silent data corruption* (*SDC*) event is recorded when the transaction completes nominally and no explicit error is reported, yet the returned data value differs from the golden reference. This distinction is particularly important because SDC reflects a communication-level failure that remains hidden from ordinary protocol-status monitoring.

The architecture shown in Fig. 7 also clarifies the role of local and external observability. The DUT-side checker provides immediate access to low-level transaction observables such as timeout status, returned data, AXI response signals, and local fault flags, whereas the supervisory board performs the authoritative final classification using an independent golden reference. This dual-layer observation strategy increases robustness against common-mode corruption and allows failure recording to continue even if the irradiated DUT experiences partial instability.

To support temporal correlation with routing-delay measurements, all physical, configuration-level, and architectural observations are stored in a synchronized cross-layer event record, as shown at the bottom of Fig. 7. This record includes RO frequency measurements, SEM status information, frame or injection identifiers (ID), AXI transaction logs, timestamps, and neutron fluence or beam-exposure information. By aligning these data streams in time, the platform enables each communication-level event to be examined together with the routing-delay state, configuration-management status, and irradiation context present before, during, and after the event. This unified logging mechanism is essential for determining whether routing-delay shifts remain localized physical disturbances or instead precede, accompany, or predict transaction-level failure modes.

Taken together, the AXI benchmark and correctness-monitoring framework provide the architectural observability that complements the physical sensing capability of the RO array. While the distributed ROs detect routing-delay perturbations in frame-localized interconnect regions, the benchmark determines whether the communication fabric continues to deliver correct transaction results. The combination of these two observation layers constitutes the central cross-layer measurement mechanism of the platform and forms the basis for the correlation and cross-section analyses presented in the subsequent sections.

### *E. Frame-Level Configuration Fault Injection Infrastructure*

In addition to neutron irradiation, the experimental platform incorporates a frame-level configuration fault injection infrastructure to provide controlled and repeatable emulation of configuration-memory upsets. Neutron exposure remains the primary experimental condition of interest because it represents the stochastic particle-induced upset environment addressed in this work. However, fault injection provides a complementary mechanism for localizing sensitive frame regions, reproducing selected perturbation scenarios, and comparing controlled configuration disturbances with neutron-induced events using the same measurement and classification framework.

The implemented injection workflow is shown in Fig. 8. Frame selection is first derived from the frame map established in Fig. 6, with emphasis on configuration frames associated with the AXI interconnect routing region and nearby RO instances. This frame-aware selection is essential because the objective is not merely to determine whether an injected upset produces a functional error, but also to evaluate whether perturbations in specific frame regions produce measurable RO frequency shifts and whether those shifts are temporally associated with AXI-level outcomes.

Controlled perturbations are introduced through the SEM-assisted configuration access infrastructure. The SEM controller is used to support configuration-event monitoring, classification, and controlled upset emulation, while the internal configuration access path (ICAP) provides frame-level access to the target configuration memory. As indicated in Fig. 8, the injector targets selected configuration frames in the DUT while the RO array and AXI benchmark remain active whenever possible. This allows the same experimental run to capture physical routing-delay response, transaction-level correctness, and configuration-management status following each injected perturbation.

For each injection, the platform records the frame identifier, injection identifier (ID), timestamp, SEM status, RO frequency response, AXI transaction outcome, and recovery or restoration status. These fields are stored in the synchronized cross-layer event record shown in Fig. 8. This record extends the logging structure introduced in Fig. 7 by adding injection-specific information, thereby allowing injected events to be analyzed on the same temporal basis as neutron-induced events. As a result, the experiment can determine whether a given frame-level perturbation produces no observable effect, a routing delay event only, an AXI-level availability failure, a detectable protocol error, or an SDC.

After each injection, the configuration state is restored or the DUT is reset to a known baseline before the next injection is applied. This recovery step, shown at the bottom of Fig. 8, preserves experimental repeatability and prevents ambiguity caused by the accumulation of multiple uncontrolled configuration changes. When an injection causes persistent instability or loss of forward progress, the failure is logged before recovery is initiated. This ensures that severe events remain part of the dataset while subsequent injections are still performed from a controlled state.

The injection infrastructure is not intended to replace beam testing. Rather, it serves three complementary roles. **First**, it provides deterministic access to selected configuration-frame regions that may not be struck during a finite neutron exposure. **Second**, it enables repeated probing of frame regions surrounding the AXI interconnect and distributed RO array, supporting spatial sensitivity mapping. **Third**, it helps distinguish severe structural configuration failures from smaller routing perturbations that may preserve logical connectivity while reducing timing margin.

By combining stochastic neutron-induced events with controlled frame-level injection, the platform supports both physical realism and experimental repeatability. Neutron irradiation exposes the operational system to representative particle-induced upset processes, while fault injection provides controlled localization and reproducibility. The resulting dual-mode methodology strengthens the subsequent correlation analysis by allowing routing-delay events, frame locations, SEM reports, AXI transaction outcomes, and recovery behavior to be interpreted within a common frame-aware experimental framework.

### *F. Neutron Irradiation Procedure and Synchronized Data Acquisition*

The objective of the neutron irradiation beam experiment is not only to observe whether upsets occur, but also to determine how naturally occurring radiation-induced perturbations propagate through the routing fabric and into AXI-level communication behavior. The overall irradiation run sequence and synchronized acquisition process are illustrated in Fig. 9. As shown in Fig. 9(a), the experiment begins with a pre-beam baseline phase ❶ in which the RO array, AXI benchmark, SEM monitor, and supervisory logging system are initialized and verified. During this phase, the nominal frequency of each RO is recorded under stable operating conditions, the AXI benchmark is executed to confirm correct transaction behavior, and the SEM controller is placed in a known monitoring state. This baseline provides the reference against which later frequency deviations, communication failures, and configuration-event reports are interpreted.

During beam exposure ❷, the irradiated DUT ZCU104 board is positioned inside the neutron beam, while the supervisory ZCU104 board remains outside the beam. The two boards remain connected through the FMC link so that transaction records, checker outputs, event markers, and synchronization information can be transferred from the DUT to the non-irradiated supervisory platform. This separation ensures that the golden-reference computation and primary logging infrastructure are not exposed to the same neutron field as the DUT, thereby reducing the likelihood of common-mode corruption in the correctness-checking path.

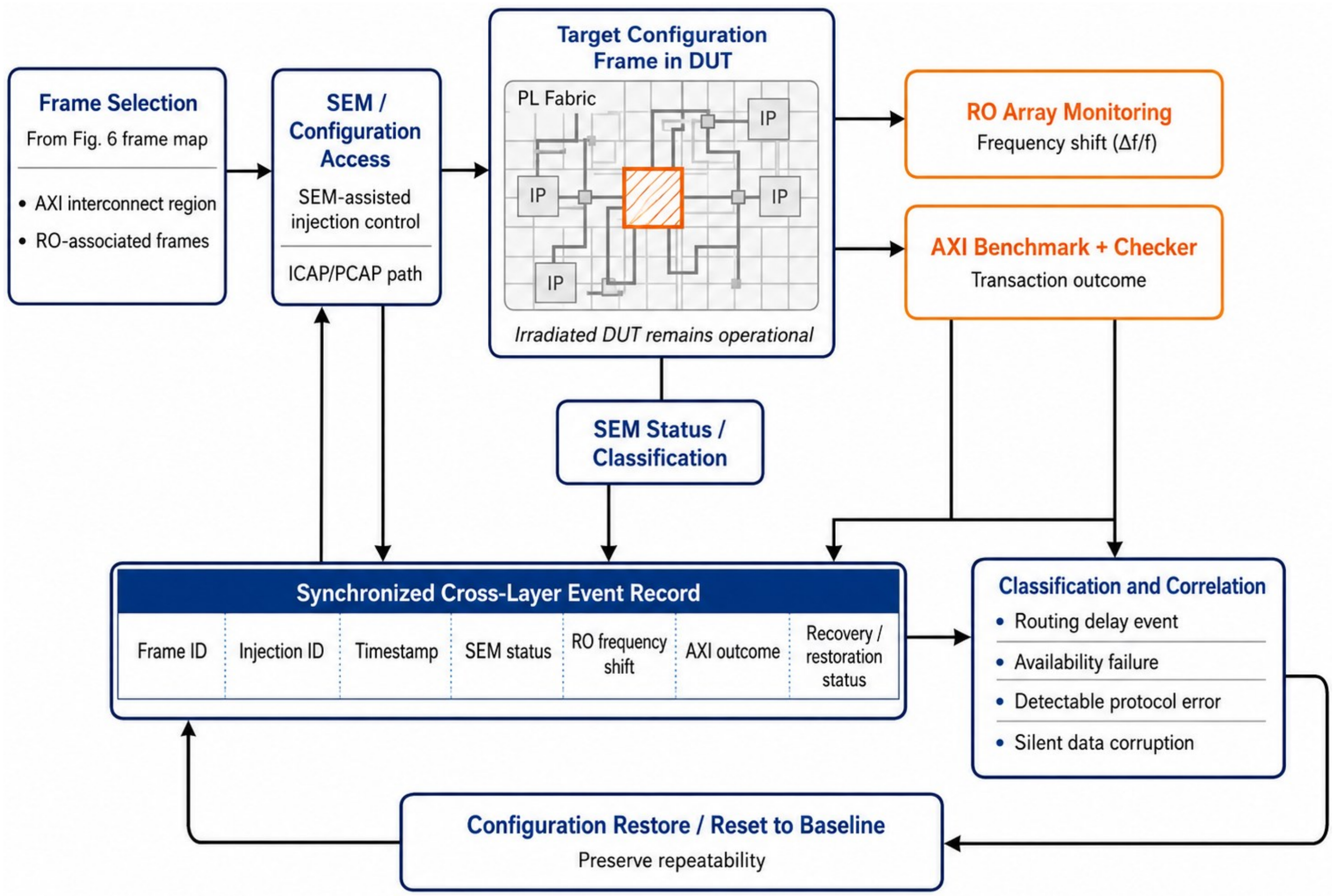

Fig. 8. Frame-level configuration fault injection and event-correlation workflow. Configuration frames associated with the AXI interconnect routing region and distributed RO array are selected using the frame map established in Fig. 6. The SEM-assisted configuration access infrastructure injects controlled perturbations into selected frames while the AXI benchmark and RO measurement system remain active. Each injection is logged with frame identifier, injection identifier, timestamp, SEM status, RO frequency response, AXI transaction outcome, and recovery status. The configuration state is restored after each injection to preserve repeatability and prevent ambiguity from accumulated faults.

Once the beam is enabled, the AXI workload, RO measurement system, SEM monitor, and supervisory acquisition process remain active continuously. The processing system issues deterministic AXI transactions through the shared interconnect to the slave-side accelerator IPs, while the RO counters periodically sample oscillator frequencies. Each RO measurement is timestamped and compared against its pre-beam baseline to identify routing-delay events. At the same time, AXI transaction outcomes are classified as correct transactions, availability failures, detectable protocol errors, or silent data corruption according to the event taxonomy defined in Section III.

As shown in the lower timeline of Fig. 9(a), all physical, architectural, configuration-level, and irradiation-related observations are stored in a synchronized event record. The logged streams include RO frequency measurements, AXI transaction outcomes, SEM status reports, neutron-fluence information, timestamps, timeout or watchdog indicators, and recovery actions. These streams are aligned to a common temporal axis so that each observed communication-level event can be examined together with the routing-delay state, SEM status, and accumulated fluence present before, during, and after the event.

When an event is detected during irradiation ❸, the system first records the observable state of the DUT and supervisory logs. If the DUT remains operational, the experiment continues without interruption so that subsequent behavior can be captured under accumulated exposure. If a persistent stall, availability failure, or loss of forward progress occurs, the final observable state is logged before a recovery or reinitialization action is applied ❹. Depending on the severity of the event, recovery may consist of restarting the AXI workload, resetting selected programmable-logic modules, invoking configuration recovery mechanisms, or reinitializing the DUT to a known baseline. Each recovery action is timestamped and included in the synchronized event record.

After beam exposure, a post-run validation phase ❺ is performed. The RO array is remeasured to determine whether observed frequency shifts have persisted or returned to baseline. The AXI benchmark is rerun to verify whether transaction correctness has been restored, and the SEM and event logs are examined to distinguish transient operational disturbances from persistent configuration-induced effects. This post-run phase provides an additional consistency check on event classification and helps identify whether recovery actions successfully returned the system to a known operational state. The neutron irradiation procedure therefore provides the physically representative counterpart to the controlled injection workflow described in Subsection IV-E. Beam exposure captures the stochastic nature of radiation-induced configuration upsets in an operational system, while synchronized acquisition preserves the timing, configuration, fluence, and transaction-level context needed for cross-layer analysis. This methodology enables routing-delay-event cross-sections, AXI failure cross-sections, and temporal correlation metrics to be extracted from a unified experimental dataset. A snapshot of the installation setup at TRIMUF laboratory is shown in Fig. 9(b) where the blue arrow is just an annotation of the beam direction toward the DUT board and is not the real neutron beam.

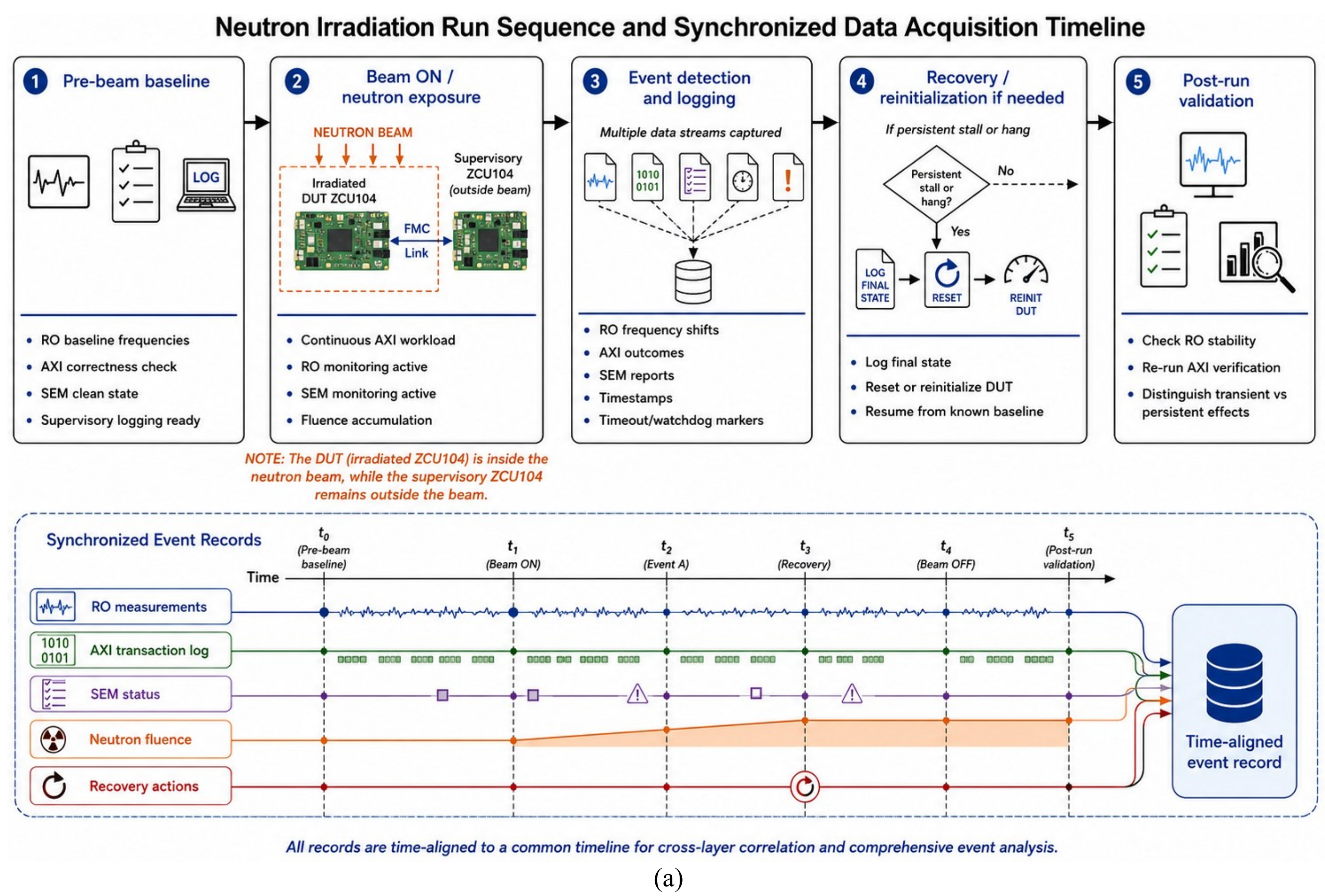

(a)

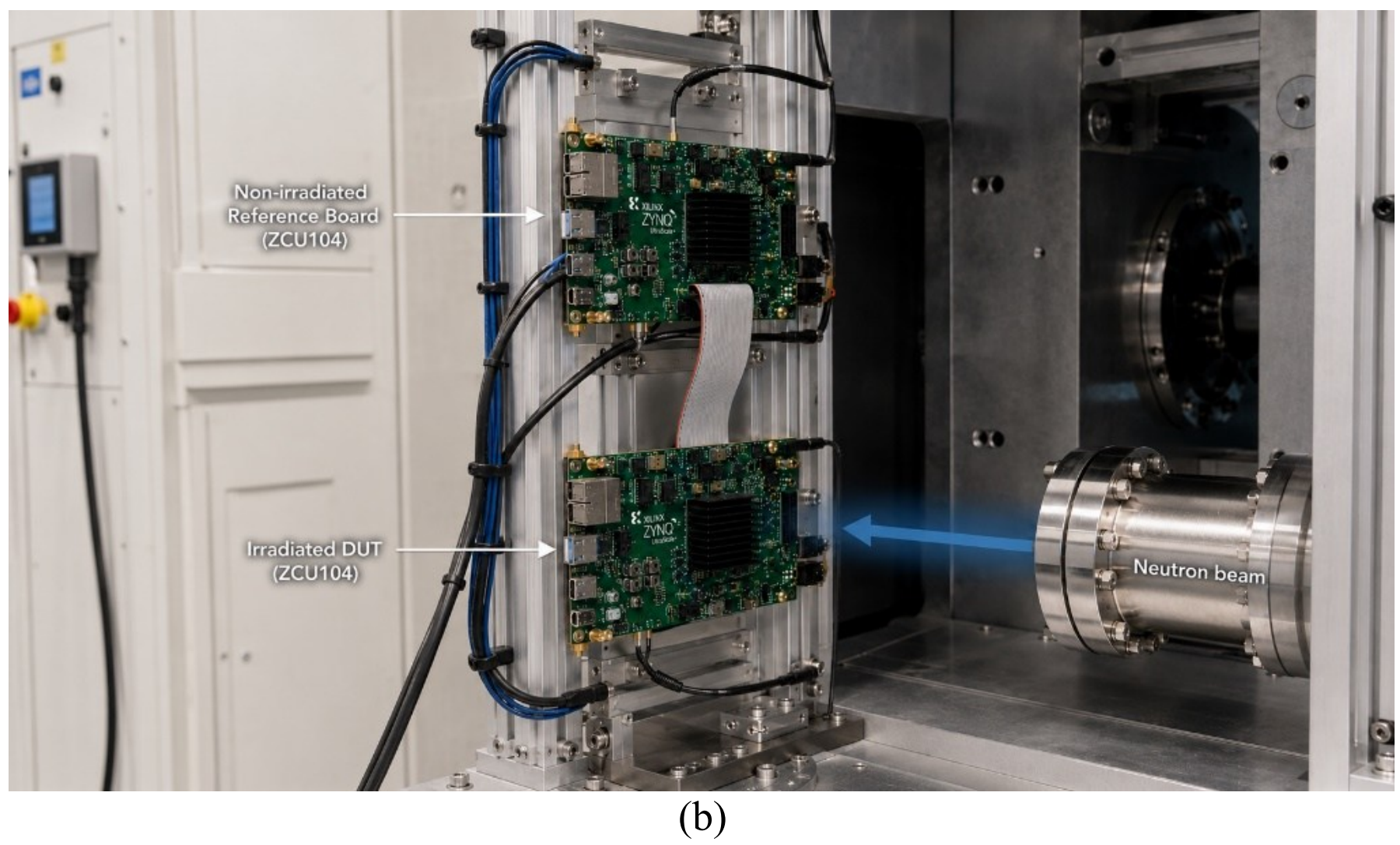

(b)

Fig. 9. Neutron irradiation setup and synchronized experimental procedure. (a) Neutron irradiation run sequence and synchronized data-acquisition timeline, including pre-beam baseline characterization, beam exposure, RO monitoring, AXI transaction logging, SEM-status recording, fluence tracking, recovery actions, and post-run validation. (b) Physical installation of the irradiated ZCU104 DUT and reference board at the neutron beam exposure position, with mechanical support, cabling, and data-acquisition connections arranged for in-beam operation.

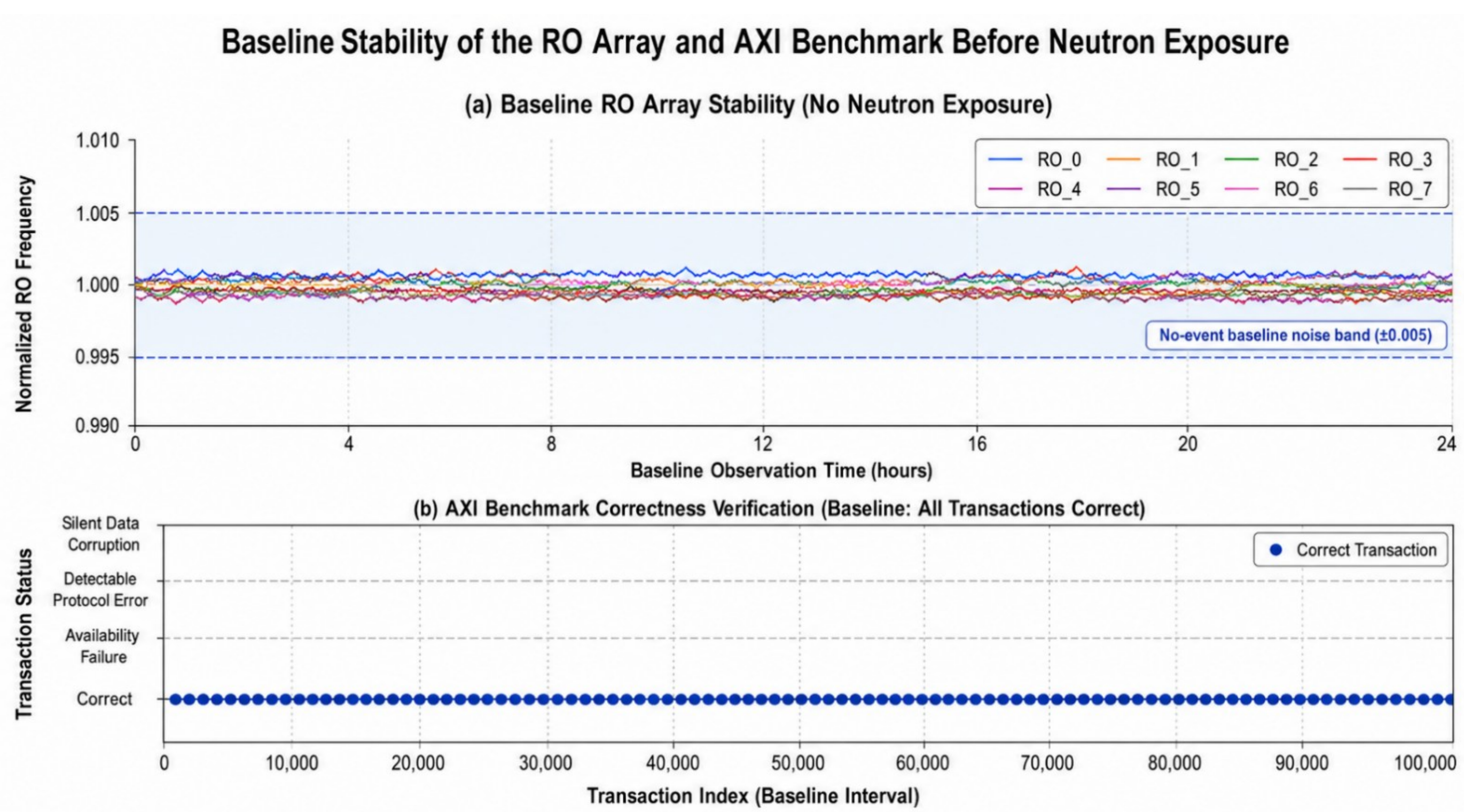

Fig. 10. Baseline stability of the routing-dominated RO array and AXI benchmark before neutron exposure. (a) Measured normalized RO frequency traces during the pre-beam baseline interval. The traces remain inside the predefined no-event noise band, indicating stable RO behavior before irradiation. (b) AXI benchmark correctness verification during the same baseline phase. All transactions are classified as correct, with no availability failure, detectable protocol error, or silent data corruption observed.

## V. EXPERIMENTAL RESULTS

This section presents the experimental results obtained from the cross-layer measurement platform described in Section IV. The analysis is organized to progress from baseline characterization to radiation-induced routing-delay observations, AXI communication failures, temporal correlation, and comparison with controlled frame-level fault injection. This structure follows the physical-to-architectural propagation chain introduced in Section III and allows routing-level and transaction-level phenomena to be evaluated within a common experimental framework.

### *A. Baseline Characterization of RO Sensors and AXI Workload*

Before neutron exposure and frame-level configuration fault injection, the implemented system was characterized under nominal operating conditions to establish reference behavior for both the routing-delay sensors and the AXI benchmark. This baseline characterization is necessary because the proposed methodology relies on detecting deviations from normal timing behavior and distinguishing radiation-induced or injection-induced events from ordinary measurement noise, environmental drift, and workload-related variability.

The baseline stability measurements of the RO array and the AXI benchmark are shown in Fig. 10. During the pre-beam interval, the routing-dominated ROs were enabled and sampled while the DUT operated under stable PVT conditions. For each RO, the measured frequency was normalized to its nominal baseline value so that frequency deviations could be compared across sensors with different absolute oscillation frequencies. As shown in Fig. 10(a), the measured RO traces remain confined within the predefined baseline noise band, indicating that no routing-delay event is detected during the nominal observation interval.

The no-event baseline noise band ($\pm 0.005$) shown in Fig. 10(a) defines the reference stability region used for later delay-event detection. Frequency deviations remaining inside this band are treated as baseline fluctuation, whereas excursions exceeding the selected threshold are candidates for routing-delay events during irradiation or injection experiments. This baseline thresholding step is essential because RO frequency can be affected by ordinary environmental and measurement variations even in the absence of radiation-induced configuration upsets. By establishing the baseline noise floor before exposure, subsequent RO frequency shifts can be interpreted with greater confidence.

In parallel with RO characterization, the AXI benchmark was executed continuously using the deterministic transaction patterns described in Subsection IV-D. The purpose of this test was to verify that the AXI interconnect, replicated slave-side IPs, DUT checker, FMC transfer path, and supervisory golden-reference classification logic operate correctly before any radiation-induced or injected perturbation is introduced. As shown in Fig. 10(b), all baseline transactions are classified as correct transactions. No availability failure, detectable protocol error, or silent data corruption is observed during the pre-beam characterization interval.

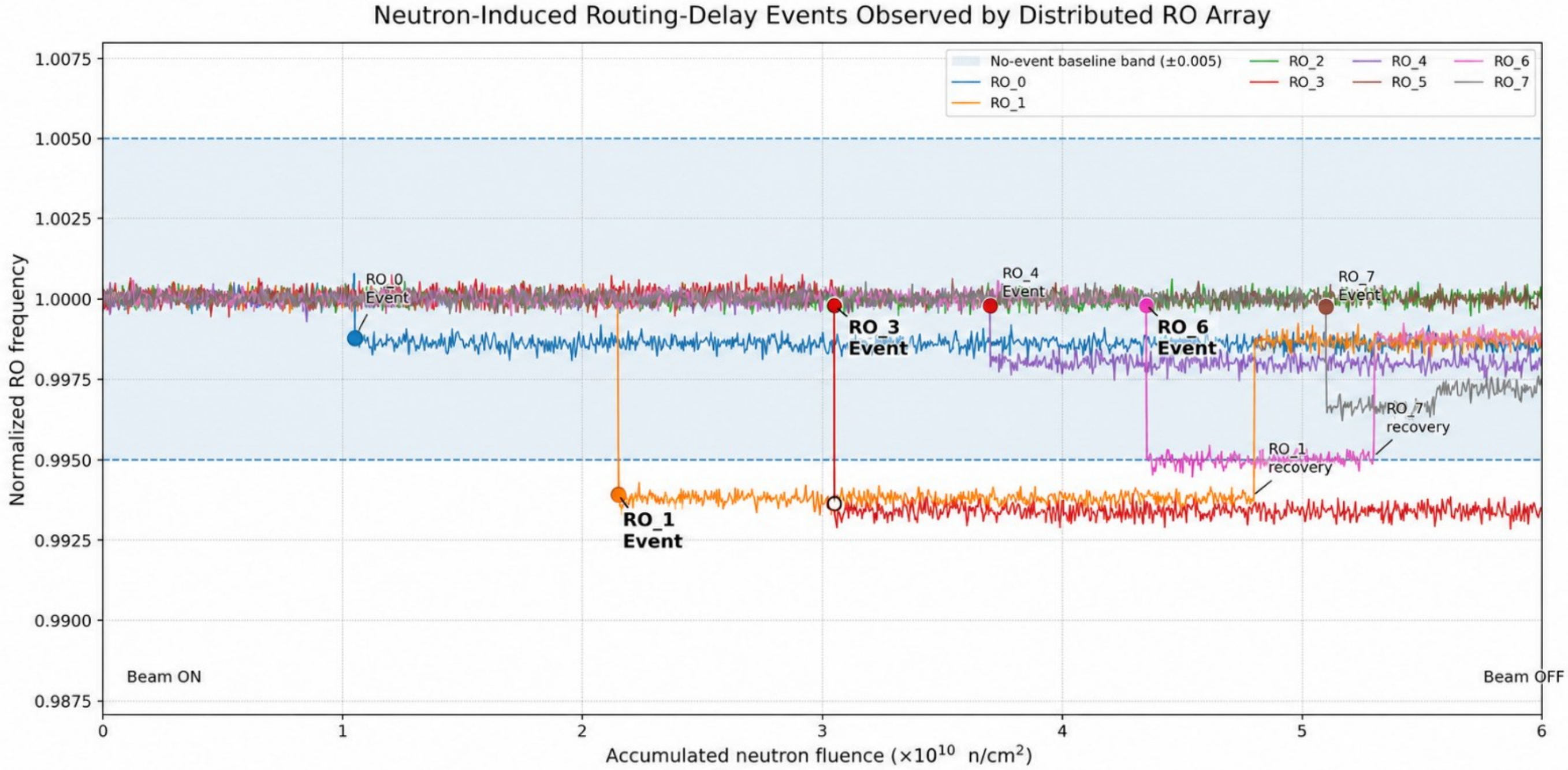

Fig. 11. Neutron-induced routing-delay events observed by the distributed RO array during beam exposure.

The SEM monitoring infrastructure was also initialized and observed during the baseline phase. No configuration-event report associated with the monitored design region is expected during this interval. This establishes a clean initial state for subsequent neutron and fault-injection experiments and ensures that later configuration-related observations are not confused with initialization artifacts or pre-existing configuration inconsistencies.

The baseline characterization therefore defines the reference state for the experimental results that follow. Routing-delay events are evaluated relative to the nominal RO behavior and threshold band established in Fig. 10(a), while AXI communication failures are identified relative to the error-free transaction behavior verified in Fig. 10(b). Together, these baseline measurements confirm that the cross-layer measurement platform is stable before exposure and that subsequent deviations can be attributed to neutron-induced or injected configuration perturbations rather than to pre-existing instability.

### B. Neutron-Induced Routing Delay Events

After baseline characterization, the DUT was exposed to neutron irradiation while the distributed RO array remained active and continuously monitored. The objective of this experiment was to determine whether neutron-induced configuration upsets produce measurable timing perturbations in routing resources associated with the AXI interconnect region. Because each RO was mapped to a localized configuration-frame region, the measured frequency response provides a frame-aware indication of routing-delay sensitivity during beam exposure.

For each RO instance, the measured frequency during irradiation was normalized to its pre-beam baseline value. Frequency deviations were then evaluated relative to the no-event threshold established in Subsection V-A. A routing-delay event was recorded when the normalized frequency deviation exceeded the predefined baseline noise band and exhibited a persistent or step-like change rather than a short isolated fluctuation. This criterion prevents ordinary measurement noise from being misclassified as a radiation-induced delay event while preserving sensitivity to configuration-induced changes in routing delay.

The measured RO frequencies during neutron exposure are shown in Fig. 11. Several RO traces remain within the baseline noise band over the irradiation interval (e.g., RO_2, RO_4, and RO_5), indicating that not all monitored frame regions experience measurable routing perturbation during the run. In contrast, selected RO instances (e.g., RO_1, RO_3, and RO_6) exhibit abrupt frequency deviations that exceed the no-event region. These deviations are interpreted as routing-delay events because they are consistent with neutron-induced configuration upsets affecting PIPs or related routing resources. The localized nature of these responses also supports the use of the RO-to-frame mapping introduced in Fig. 6 for identifying sensitive routing regions within the AXI interconnect fabric.

The sign of the observed frequency shifts provides information about the direction of the delay perturbation. In Fig. 11, the dominant event signatures appear as decreases in normalized RO frequency. Since the oscillation frequency of an RO is inversely related to its loop delay, a frequency decrease corresponds to an increase in effective propagation delay. This behavior is consistent with the activation of unintended parasitic routing branches, additional capacitive loading, or other SEU-induced modifications of the routing path that increase the effective delay without necessarily producing immediate logical disconnection.

The persistence of the frequency shifts is also important. Some RO traces exhibit step-like offsets that remain after the initial event, suggesting that the underlying configuration state has been modified and that the timing perturbation persists until correction or reconfiguration. Other traces show partial recovery after a later recovery or correction marker, indicating that the measured delay perturbation can be reduced when the affected configuration state or system state is restored. These behaviors motivate logging RO measurements together with SEM status, recovery markers, and AXI transaction outcomes, as described in Subsections IV-D through IV-F.

The event distribution in Fig. 11 further shows that routing-delay sensitivity is not uniform across all monitored RO locations. Some RO instances remain stable, while others exhibit larger or more persistent deviations. This nonuniform response is expected because different ROs traverse different routing resources and are associated with different configuration-frame regions. Consequently, the distributed RO array provides not only event detection but also spatial discrimination of sensitive routing regions.

The detected routing-delay events form the physical-layer event set used in the subsequent cross-layer analysis. Their timestamps and associated RO identifiers are compared with AXI communication outcomes to determine whether timing perturbations in the monitored routing region precede, accompany, or remain independent of architectural-level failures. The corresponding event counts are later normalized by accumulated neutron fluence to derive routing-delay-event cross-sections and to compare physical-layer susceptibility with AXI-level failure susceptibility.

### C. AXI Communication Failure Observations

While the distributed RO array captures physical-layer routing-delay perturbations, the AXI benchmark provides the corresponding architectural-level observation of communication correctness. During neutron exposure, the processing system continuously issued deterministic AXI transactions through the shared interconnect to the replicated slave-side accelerator IPs. Each returned value, AXI response condition, timeout state, and checker output was logged and compared against the golden reference maintained by the non-irradiated supervisory board.

The measured AXI communication outcomes during irradiation are summarized in Fig. 12. The upper panel presents the complete event distribution on a logarithmic scale, including both correct transactions and failure events. A total of 2,480,000 transactions completed correctly, confirming that the benchmark remained operational for the dominant portion of the irradiation interval. In parallel, the monitoring framework recorded 21 availability failures, 14 detectable protocol errors, and 18 SDC events. Because the correct-transaction count is several orders of magnitude larger than the failure counts, Fig. 12(b) provides an expanded linear-scale view of the failure categories.

The event classification follows the taxonomy defined in Subsection III-D and implemented by the transaction-level monitoring framework described in Subsection IV-D. Transactions completing within the timeout interval, returning the expected data value, and producing no explicit protocol error are classified as "*Correct*". Transactions failing to complete within the predefined timeout interval, or producing persistent loss of forward progress, are classified as "*Availability failures*". Transactions completing with an explicit AXI error response or another directly observable protocol-level fault indication are classified as "*Detectable protocol errors*". Transactions completing without timeout or explicit protocol error but returning data inconsistent with the golden reference are classified as "*Silent data corruption*".

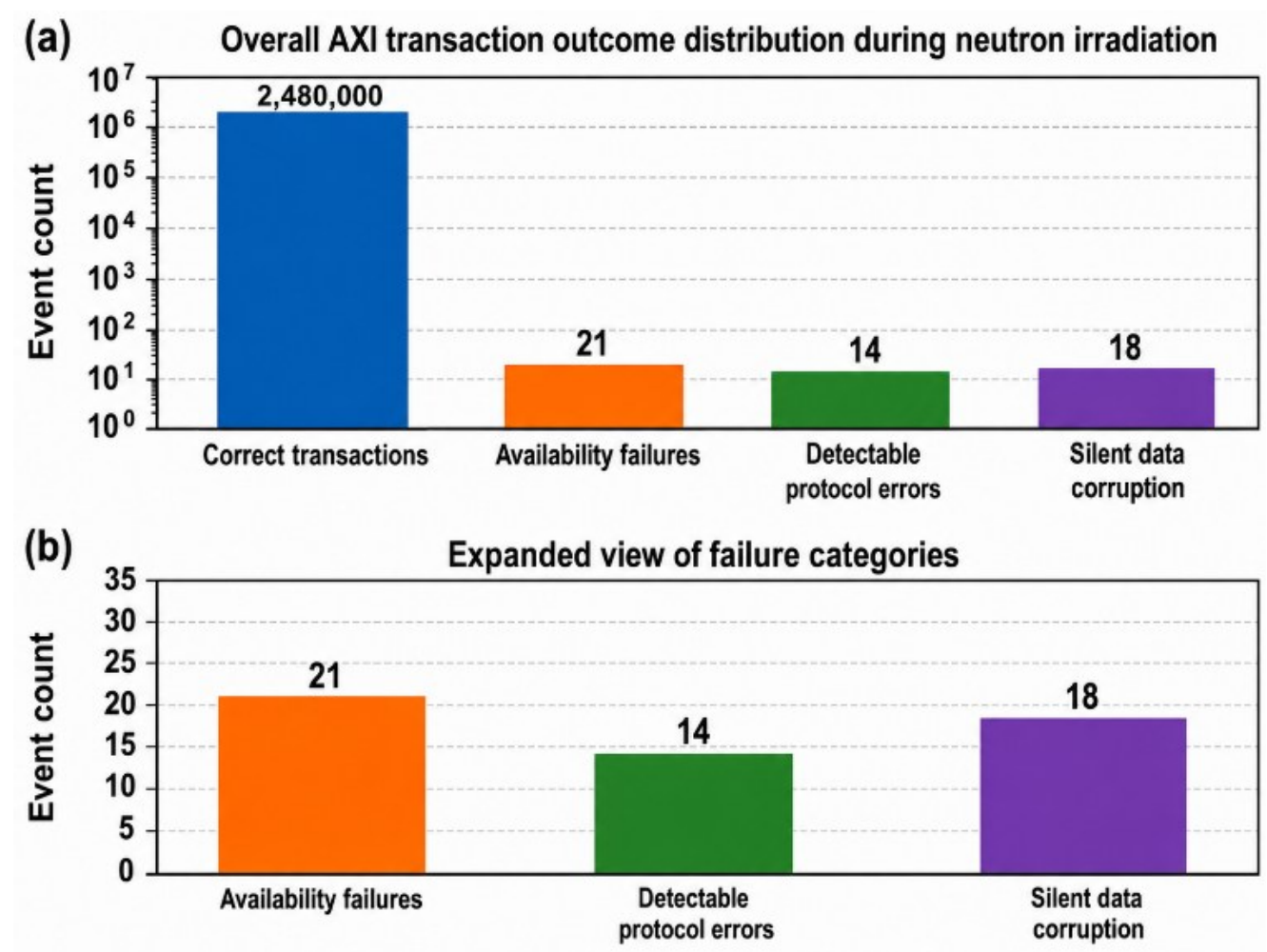

Fig. 12. Measured AXI communication outcomes during neutron irradiation. (a) Overall transaction outcome distribution, including correct transactions and failure categories, shown on a logarithmic event-count scale. The benchmark completed 2,480,000 correct transactions while 21 availability failures, 14 detectable protocol errors, and 18 silent data corruption events were recorded. (b) Expanded linear-scale view of the failure categories. Silent data corruption corresponds to completed transactions that returned incorrect data without timeout or explicit protocol-level error indication.

The measured availability failures represent the most externally visible AXI malfunction category. In the context of neutron-induced configuration upsets, such failures are consistent with severe disruption of routing resources, arbitration logic, address decoding, or control paths that prevent transactions from completing normally. Because these failures interrupt communication, they are directly observable through timeout or watchdog mechanisms.

Detectable protocol errors form a separate class of communication failure. In these cases, the transaction does not complete as a correct transfer, but the abnormal condition is explicitly indicated through a protocol-level response or error flag. Although these events still represent architectural malfunction, their detectability allows them to be distinguished from silent corruption and potentially handled by conventional recovery mechanisms such as transaction retry, interrupt handling, or system reinitialization.

SDC is the most critical communication-level outcome observed in Fig. 12. The measured 18 SDC events correspond to completed transactions for which the returned value differed from the golden reference while no timeout or explicit protocol-level error was reported. Such events are particularly important because they can propagate incorrect data to higher software or system layers as if the transaction were valid. This behavior is consistent with the timing-margin erosion mechanism introduced in Subsection III-B, where routing-induced delay perturbations may disturb data or control sampling without necessarily producing a complete interconnect disconnection.

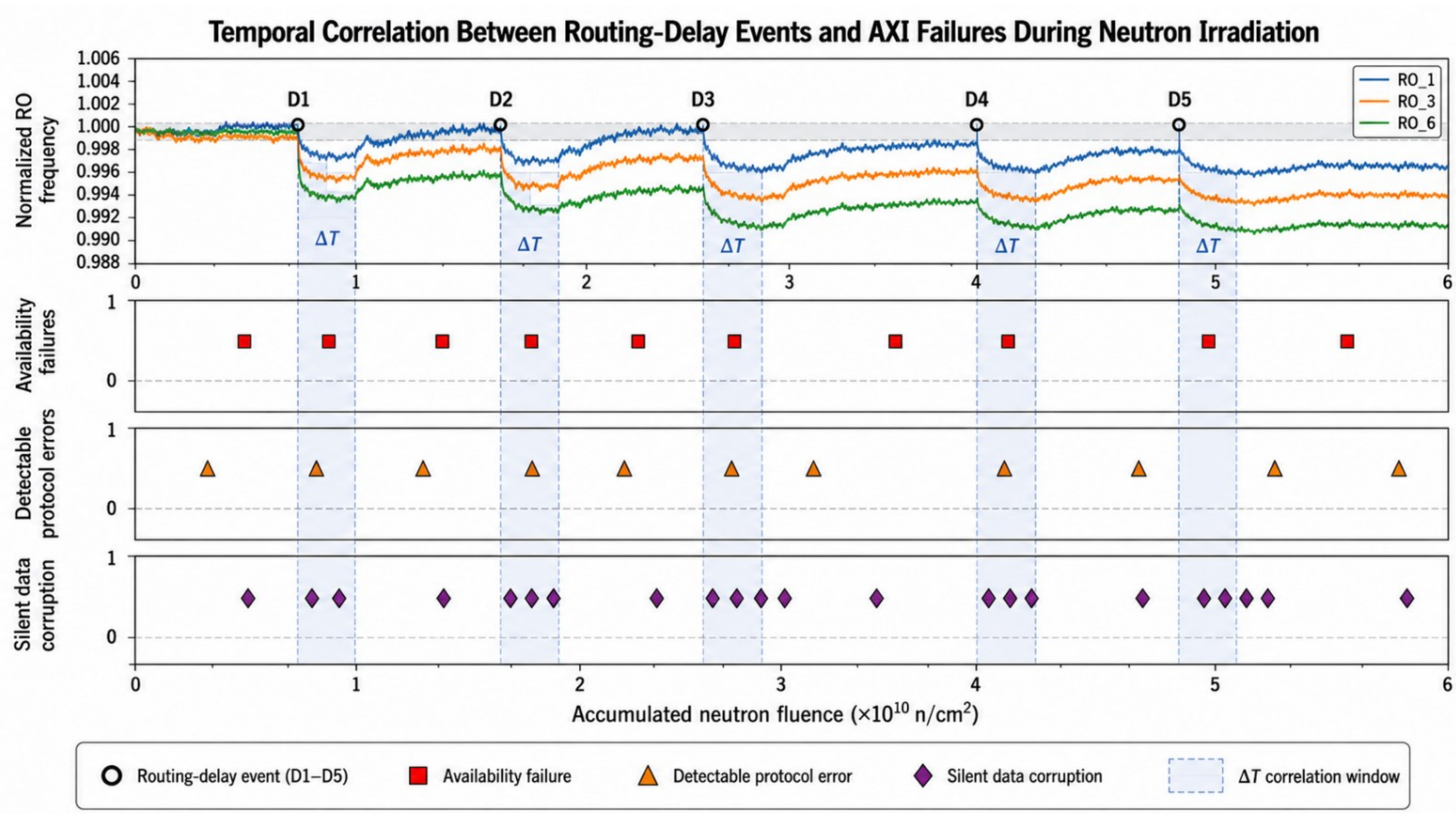

Fig. 13. Temporal correlation between routing-delay events and AXI communication failures during neutron irradiation. Selected RO frequency traces are shown for monitored frame regions in which routing-delay events were detected and used for correlation analysis; the complete RO-array response is shown in Fig. 11.

The distribution in Fig. 12 therefore demonstrates that neutron-induced perturbations can produce multiple architectural manifestations in the AXI communication fabric. Availability failures indicate loss of service, detectable protocol errors indicate signaled communication malfunction, and SDC indicates undetected functional corruption. These measured AXI-level events form the architectural event set used in the next subsection, where they are time-aligned with routing-delay events detected by the RO array to evaluate cross-layer correlation.

### *D. Temporal Correlation Between Routing Delay Events and AXI Failures*

The results presented in Subsections V-B and V-C show that neutron irradiation produces both routing-level delay perturbations and AXI-level communication failures. The central question of this work is "*whether these two event classes occur independently or whether routing-delay events provide observable precursors or correlates of architectural failure*". To address this question, the timestamped routing-delay events detected by the distributed RO array were time-aligned with the AXI transaction outcomes recorded by the supervisory monitoring framework.

For each detected routing-delay event, a temporal association window of duration $\Delta T$ was defined, as introduced in Section III. AXI failures occurring within this interval after a routing-delay event were classified as temporally associated events. The analysis was performed separately for availability failures, detectable protocol errors, and SDC because these failure modes represent different architectural manifestations. This separation is important because a severe configuration disruption may produce an immediate timeout or stall, whereas a smaller delay perturbation may allow transactions to complete while still corrupting sampled data. The time-aligned event analysis is illustrated in Fig. 13. For clarity, the upper panel displays selected RO traces associated with monitored frame regions derived from Fig. 11 in which routing-delay events were detected and used for the correlation analysis.

In Fig. 13, routing-delay events are labeled $D1$ through $D5$ at the fluence values where selected RO traces exhibit abrupt frequency shifts. The shaded regions following these events denote the temporal association windows $\Delta T$. The lower panels show availability failures, detectable protocol errors, and SDC events recorded by the AXI benchmark over the same irradiation interval. By plotting all event classes against a common accumulated-fluence axis, the figure allows the temporal proximity between physical routing perturbations and architectural communication failures to be examined directly.

The correlation timeline shows that a subset of AXI communication failures occurs within the $\Delta T$ windows following routing-delay events. SDC events are observed near several RO frequency-shift transitions, indicating that routing perturbations can coincide with communication-level corruption without necessarily producing a hard interconnect failure. This behavior is consistent with the timing-margin erosion mechanism described in Subsection III-B. In contrast, availability failures appearing near larger or more abrupt routing perturbations are consistent with more severe disruption of routing, arbitration, or control paths within the shared AXI interconnect fabric.

Not all routing-delay events lead to observable AXI failure. Some RO frequency shifts occur without a corresponding transaction error within the selected association window. This outcome is expected because an RO may detect a perturbation in a monitored frame region that does not lie on an active or timing-critical AXI path during the observation interval. Similarly, not every AXI failure is necessarily preceded by a detected RO event, because the distributed RO array provides spatial sampling of the interconnect routing fabric rather than exhaustive coverage of all configuration-controlled routing resources. These cases emphasize that the RO array provides localized observability and statistical correlation evidence, not complete deterministic fault coverage.

The conditional failure probability $\hat{P}(F|D)$ was computed by dividing the number of AXI failures occurring within the association window after routing-delay events by the total number of detected routing-delay events. This metric was evaluated separately for availability failures, detectable protocol errors, and SDC. The resulting values quantify how often an observed routing perturbation is followed by a given architectural failure class and provide a statistical bridge between physical-layer timing behavior and system-level communication correctness.

The temporal correlation results therefore support the interpretation that routing-delay events can act as measurable indicators of AXI vulnerability under neutron irradiation. The analysis does not imply that every routing-delay event causes a communication failure, nor that every AXI failure must be preceded by a detected RO event. Instead, it demonstrates that timing perturbations observed in selected interconnect frame regions are temporally associated with a subset of availability failures and SDC events. This connection provides the cross-layer evidence required to relate configuration-level routing disturbances to architectural-level degradation in the operational MPSoC system.

### *E. Cross-Section Extraction and Conditional Correlation Metrics*

The event observations presented in Subsections V-B through V-D were converted into fluence-normalized susceptibility metrics using the statistical framework introduced in Subsection III-E. For each event class, the experimental cross-section was obtained by normalizing the number of observed events by the accumulated neutron fluence during the corresponding irradiation interval. In this experiment, the accumulated fluence used for normalization was ($6.0 \times 10^{10}\, n/cm^2$). This normalization allows routing-delay events and AXI communication failures to be compared on a common radiation-exposure basis, even though they correspond to different abstraction levels of the system.

The routing-delay-event cross-section was computed from the total number of detected RO frequency-shift events exceeding the baseline threshold established in Subsection V-A. As shown in Fig. 11, eight routing-delay events were identified across the distributed RO array during neutron exposure. This produces a routing-delay-event cross-section of ($1.33 \times 10^{-10} cm^2$). Because the RO instances are mapped to localized frame regions within the AXI interconnect fabric, this value represents the susceptibility of the monitored routing region to measurable neutron-induced timing perturbations.

The AXI-level failure cross-sections were computed separately for availability failures, detectable protocol errors, and SDC. This separation is necessary because the three event categories correspond to different architectural consequences. From the measured AXI outcomes shown in Fig. 12, 21 availability failures, 14 detectable protocol errors, and 18 SDC events were recorded. These counts correspond to cross-sections of $3.50 \times 10^{-10} cm^2$, $2.33 \times 10^{-10} cm^2$, and $3.00 \times 10^{-10} cm^2$, respectively.

The conditional correlation metrics were computed from the time-aligned event analysis shown in Fig. 13. For each detected routing-delay event ($D$), AXI failures occurring within the temporal association window $\Delta T$ were counted as associated events. The estimated conditional probability $\hat{P}(F|D)$ was then evaluated separately for each AXI failure class. Within the selected association windows, three availability failures, two detectable protocol errors, and four SDC events were temporally associated with routing-delay events. These values correspond to conditional metrics of 0.375, 0.250, and 0.500, respectively.

The extracted cross-sections and conditional metrics are summarized in Table III. The cross-section values quantify how often each event category occurs per unit neutron fluence, whereas the conditional metrics quantify how frequently a detected routing-delay event is followed by a specific AXI failure class within the correlation window. For the individual AXI failure classes, the reported conditional metrics are interpreted as estimated conditional probabilities, $\hat{P}(F_k|D)$, where $F_k$ denotes availability failure, detectable protocol error, or silent data corruption. The aggregate AXI-failure row is treated differently: the quantity $N_{F\cap D}/N_D$ is reported instead of $\hat{P}(F|D)$, because it represents the number of temporally associated AXI failure events per detected routing-delay event rather than a strict probability. This aggregate rate can exceed unity when more than one AXI failure occurs within the association windows following a single routing-delay event. The silent data corruption conditional metric is particularly important because it indicates that a significant fraction of detected routing-delay events are followed by completed but incorrect AXI transactions, supporting the timing-margin erosion mechanism

Table III. Measured Event Counts, Cross-Section, and Conditional Metrics

| Fluence $\Phi(n/cm^2) = (6.0 \times 10^{10})$ | | | | | |
|---|---|---|---|---|---|
| Cross-Section $\sigma(cm^2)$ | | | | | |
| $\sigma_D = \frac{N_D}{\Phi}$ | $\sigma_{F_A} = \frac{N_{F_A}}{\Phi}$ | | $\sigma_{F_P} = \frac{N_{F_P}}{\Phi}$ | $\sigma_{F_{SDC}} = \frac{N_{F_{SDC}}}{\Phi}$ | $\sigma_{(F_A \cup F_P \cup F_{SDC})} = \frac{N_{F_P}}{\Phi}$ |
| **Category** | **Symbol** | **Count ($N_{Symbol}$)** | **Associated Count Within $\Delta T$** | **Cross-Section** | **Conditional Metric** |
| **Routing-delay event** | $D$ | 8 | --- | $(1.33 \times 10^{-10} cm^2)$ | --- |
| **Availability failure** | $F_A$ | 21 | 3 | $(3.50 \times 10^{-10} cm^2)$ | $\hat{P}(F_A \mid D) = 0.375$ |
| **Detectable protocol error** | $F_P$ | 14 | 2 | $2.33 \times 10^{-10} cm^2$ | $\hat{P}(F_P \mid D) = 0.375$ |
| **Silent data corruption** | $F_{SDC}$ | 18 | 4 | $3.00 \times 10^{-10} cm^2$ | $\hat{P}(F_{SDC} \mid D) = 0.500$ |
| **All AXI failure events** | $F_A \cup F_P \cup F_{SDC}$ | 53 | 9 | $8.83 \times 10^{-10} cm^2$ | $N_{F \cap D}/N_D = 1.125$ |

discussed in Section III-B.

*F. Comparison with Frame-Level Fault Injection*

The neutron irradiation results presented in the previous subsections demonstrate that routing-delay events and AXI communication failures can both occur during beam exposure. However, neutron-induced upsets occur stochastically and cannot be directed to a specific configuration frame during irradiation. To complement the beam results, the frame-level fault injection infrastructure described in Subsection IV-E was used to perturb selected configuration-frame regions associated with the AXI interconnect and distributed RO array. This controlled campaign provides additional insight into the spatial sensitivity of the monitored routing region and helps interpret the relationship between configuration-frame location, RO frequency response, AXI-level outcome, and recovery behavior.

The comparison between neutron-induced responses and frame-level injected responses is summarized in Fig. 14. Each row corresponds to a monitored configuration region associated with one RO instance in the AXI interconnect routing area. The table reports whether a neutron-induced RO delay event was observed in that region, whether controlled injection reproduced an RO frequency shift, the AXI outcome observed during the injection campaign, and the corresponding recovery behavior. This format allows the beam and injection results to be compared on a common frame-localized basis.

The results show that the monitored regions do not respond uniformly. Some frame regions, such as those associated with RO2 and RO5, show no neutron-induced RO delay event and no injection-induced RO shift, while AXI transactions remain correct and no recovery action is required. These regions therefore behave as low-response regions under the tested workload and perturbation conditions. In contrast, regions associated with RO1, RO3, RO6, and RO7 (slightly) show both neutron-observed routing-delay activity and injection-induced RO shifts, indicating that these frame regions are more sensitive to configuration perturbation.

The AXI outcomes further distinguish between purely physical routing perturbations and system-level communication degradation. For some regions, such as RO0 and RO4, injected perturbations produce measurable RO shifts while the AXI benchmark remains correct. These cases indicate that a configuration disturbance can alter routing delay without immediately propagating into a transaction-level failure. This behavior is consistent with the interpretation that not every measurable routing perturbation affects an active or timing-critical AXI path during the observation interval.

Other regions exhibit both routing-delay response and AXI-level degradation. In Fig. 14, the region associated with RO3 produces a large neutron-observed delay event, an injection-induced RO shift, and an availability failure requiring reset. This behavior is consistent with a severe perturbation affecting routing or control resources that are essential for forward progress through the shared AXI interconnect. By contrast, the regions associated with RO1 and RO6 produce silent data corruption during injection and recover after correction or reinitialization. These cases are consistent with perturbations that preserve apparent transaction completion while corrupting data values, supporting the timing-margin erosion mechanism discussed in Subsection III-B.

The recovery behavior provides additional information about the severity of each perturbation. "*Persistent offsets*" indicate that the routing-delay change remains observable after

**Neutron vs. Frame-Level Fault Injection Across Monitored Configuration Regions**

| Frame Region / RO ID | Neutron RO Delay Event | Injection RO Shift | Injection AXI Outcome | Recovery Behavior |
|---|---|---|---|---|
| RO0 | Yes (small) | Yes | Correct | Persistent offset |
| RO1 | Yes | Yes | Silent data corruption | Recovered |
| RO2 | No | No | Correct | No action |
| RO3 | Yes (large) | Yes | Availability failure | Reset required |
| RO4 | Yes | Yes | Correct | Recovered |
| RO5 | No | No | Correct | No action |
| RO6 | Yes | Yes | Silent data corruption | Recovered |
| RO7 | Yes | Yes | Detectable protocol error | Recovered |

green = correct / no action; yellow = delay shift / persistent offset; orange = SDC or detectable protocol error; red = availability failure / reset required; gray = no event / no shift

*Only a subset of frame regions produces both measurable routing-delay perturbation and AXI-level degradation.*

Fig. 14. Neutron-induced and frame-level injected responses across monitored configuration regions. Each row corresponds to a configuration-frame region associated with one distributed RO instance in the AXI interconnect routing area.

the initial event and may require configuration restoration to return fully to baseline. "*Recovered*" cases indicate that correction or reinitialization reduces the observed effect and restores normal transaction behavior. "*Reset-required*" cases indicate more severe architectural disruption, such as persistent loss of forward progress or interconnect nonresponsiveness. Therefore, the recovery behavior column in Fig. 14 helps distinguish benign delay shifts from perturbations that produce more severe system-level consequences.

The injection campaign complements the neutron irradiation results in two important ways. *First*, it provides repeatable access to selected frame regions that may be struck only rarely during finite beam exposure. *Second*, it separates structural sensitivity from stochastic exposure probability. A frame region that repeatedly produces RO shifts or AXI failures under injection can be interpreted as structurally sensitive, whereas neutron irradiation determines how such sensitivity manifests under realistic particle exposure. The agreement between neutron-observed delay activity and injection-induced responses in several monitored regions strengthens the frame-localized interpretation of routing-induced communication failures.

Overall, Fig. 14 supports the conclusion that only a subset of configuration-frame regions produces both measurable routing-delay perturbation and AXI-level degradation. This result is important for mitigation because it indicates that vulnerability is not uniformly distributed across the AXI interconnect routing fabric. Instead, sensitivity depends on the relationship between configuration-frame location, active routing resources, timing-critical communication paths, and recovery behavior. This observation provides a practical basis for mitigation strategies based on placement constraints, routing isolation, selective scrubbing, and frame-aware monitoring of the most sensitive interconnect regions.

## VI. DISCUSSION

The experimental results presented in Section V demonstrate that neutron-induced effects in AXI-based FPGA SoCs cannot be fully understood from either routing-level timing measurements or transaction-level failure observations alone. This section interprets the results from a cross-layer perspective, focusing on how configuration-induced routing perturbations may propagate into architectural communication failures, why SDC is a particularly important outcome, and what the findings imply for mitigation and future system design.

### *A. Physical-to-Architectural Fault Propagation*

The results presented in Section V support the central premise of this work: *neutron-induced configuration upsets can produce measurable routing-delay perturbations that are observable at the physical layer* and, in a subset of cases, *temporally associated with AXI-level communication failures*. This observation connects two classes of effects that have often been studied separately: routing-level timing degradation and architectural-level transaction malfunction. The distributed RO array provides direct evidence of localized timing perturbation, while the AXI benchmark demonstrates whether the communication fabric continues to operate correctly under the same exposure conditions.

The correlation results do not imply that every routing-delay event necessarily causes a system-level failure. Several delay events occur without an associated AXI malfunction, which is expected because the affected routing resources may not belong to an active AXI path, may not be timing-critical during the exercised workload, or may introduce delays smaller than the available timing margin. Conversely, some AXI failures are not preceded by a detected RO event, reflecting the fact that the RO array samples selected frame-localized routing regions rather than all configuration-controlled resources in the interconnect fabric. These observations emphasize that the proposed platform provides spatially localized and statistically interpretable observability rather than exhaustive fault coverage.

Nevertheless, the measured temporal association between selected routing-delay events and AXI failures indicates that routing perturbations can act as useful indicators of communication vulnerability. In particular, delay events occurring in frame regions associated with shared interconnect routing are more relevant to architectural degradation than isolated perturbations in unused or noncritical fabric resources. This frame-localized interpretation is strengthened by the fault-injection comparison, which shows that selected configuration regions can reproduce both RO frequency shifts and AXI-level failure modes under controlled perturbation.

### *B. Interpretation of Silent Data Corruption*

SDC is the most significant failure mode observed in this study because it represents an architectural error that is not accompanied by an explicit timeout or protocol-level error response. Unlike availability failures, which interrupt forward progress and are therefore relatively easy to detect, SDC events allow transactions to be completed while delivering incorrect data. This behavior is particularly concerning in AXI-based MPSoC systems because corrupted data may propagate to software or downstream hardware as if it were valid.

The observed SDC events are consistent with the timing-margin erosion mechanism developed in Subsection III-B. A routing perturbation that increases propagation delay does not necessarily disconnect the interconnection or violate AXI protocol sequencing at a visible level. Instead, it may reduce setup margin on a data, control, or handshake path. If the affected signal is sampled incorrectly, the transaction may still complete but return an incorrect value. This mechanism provides a plausible physical explanation for why some perturbations lead to SDC rather than availability failure.

The distinction between SDC and hard failure is important for reliability assessment. A design that appears robust when evaluated only through timeout or watchdog mechanisms may still be vulnerable to undetected data corruption. Therefore, radiation testing of AXI-based systems should include independent golden-reference checking of transaction values, not merely observation of system hangs, reset events, or protocol error flags. The dual-board architecture used in this work directly addresses this requirement by placing the golden reference outside the neutron field.

### *C. Mitigation Implications*

The results suggest that mitigation should not be limited to correcting configuration memory after an error is detected. While scrubbing and SEM-based recovery are essential, they may not prevent short-lived timing-margin erosion or transient SDC events that occur before correction completes. A complete mitigation strategy should therefore combine configuration recovery with architectural and physical design measures that reduce the probability that routing perturbations propagate into transaction-level failures.

Frame-aware placement and routing constraints provide one promising direction. Since the results show that sensitivity is not uniformly distributed across monitored regions, critical AXI paths and arbitration logic should be placed and routed to reduce concentration of vulnerable shared resources. Isolation between replicated accelerators may also reduce the probability that a single routing perturbation affects multiple communication paths simultaneously. The comparison with frame-level fault injection further suggests that mitigation effectiveness should be evaluated not only by whether a fault produces a functional error, but also by whether it produces measurable timing perturbation in sensitive interconnect regions.

Runtime monitoring can also improve resilience. Distributed RO sensors placed near critical communication fabrics provide a low-level indication of routing-delay perturbation that may occur before or during architectural malfunction. Such monitors could be used to trigger increased scrubbing frequency, transaction replay, selective reinitialization, or temporary isolation of affected accelerators. However, because not every RO shift leads to an AXI failure, monitor outputs should be interpreted probabilistically and combined with transaction-level checking rather than used as standalone failure indicators.

### *D. Scope and Limitations*

The proposed methodology intentionally focuses on the AXI interconnect region of a Zynq UltraScale+ MPSoC implemented on the ZCU104 platform. The conclusions are therefore most directly applicable to designs with similar shared AXI communication structures, routing-resource concentration, and configuration-memory organization. Other FPGA families with different interconnection topologies or workloads may exhibit different sensitivity distributions and different failure-mode proportions.

The RO array provides localized observability of routing-delay perturbations, but it does not monitor every routing resource in the device. Absence of an RO event before an AXI failure only indicates that no perturbation was detected by the monitored RO locations. Increasing RO spatial coverage would improve observability, but it would also introduce additional resource overhead and could perturb the placement and routing of the benchmark itself. A practical implementation must therefore balance observability against intrusiveness.

The fault-injection campaign complements neutron irradiation but is not equivalent to beam exposure. Injection provides deterministic access to selected configuration frames and supports repeatable sensitivity studies, whereas neutron irradiation produces stochastic particle-induced events with realistic spatial and temporal distributions. Agreement between injection and beam observations strengthens interpretation, but injection cannot fully reproduce the physical charge-deposition process or all multi-bit upset patterns. For this reason, the combined beam-and-injection methodology is more informative than either approach alone.

Finally, the reported conditional correlation metrics should be interpreted as statistical association measures rather than deterministic causality proofs. The value of the proposed framework lies in providing simultaneous physical, configuration-level, and architectural observability, thereby making such cross-layer associations measurable and quantitatively analyzable.

Because the neutron irradiation campaign was conducted under a self-funded budget and without external programmatic support, the available beam time was necessarily limited. The resulting dataset is therefore interpreted together with the frame-level fault-injection campaign, which provides repeatable access to selected configuration-frame regions and complements the stochastic beam observations. Additional beam campaigns with longer exposure time and repeated runs are identified as future work to further improve statistical confidence.

## VII. CONCLUSION AND FUTURE WORK

This paper presented a cross-layer experimental study of neutron-induced configuration upsets in AXI-based Zynq UltraScale+ MPSoCs. The work addressed the unresolved connection between routing-level timing degradation and architectural-level communication failure by combining distributed routing-dominated RO sensors, AXI transaction-level correctness monitoring, SEM-assisted frame-level fault injection, and neutron irradiation of a fully operational ZCU104-based system.

The experimental platform enabled simultaneous observation of physical routing-delay perturbations and AXI communication outcomes. The distributed RO array provided localized timing observability within the AXI interconnect routing region, while the dual-board architecture allowed transaction correctness to be verified against an external golden reference outside the neutron field. This arrangement made it possible to distinguish availability failures, detectable protocol errors, and silent data corruption during irradiation.

The results showed that neutron exposure can produce measurable routing-delay events in selected configuration-frame regions and that a subset of these events is temporally associated with AXI-level communication failures. In particular, silent data corruption events were observed in which AXI transactions completed without explicit protocol-level error while returning incorrect data. This behavior supports the interpretation that routing-induced timing-margin erosion may lead to architectural failure without necessarily producing a hard interconnect disconnection or system stall.

Frame-level fault injection further supported the frame-localized interpretation of the results. Controlled perturbations in selected configuration regions reproduced several qualitative behaviors observed under neutron exposure, including RO frequency shifts, correct-but-delayed operation, silent data corruption, availability failure, and recovery-dependent behavior. These observations indicate that vulnerability is not uniformly distributed across the interconnection fabric but depends on the relationship between configuration-frame location, active routing resources, timing-critical communication paths, and recovery mechanisms.

Combining beam testing with frame-aware fault injection and synchronized cross-layer logging enables routing-delay-event cross-sections, AXI failure cross-sections, and conditional correlation metrics to be extracted from a unified dataset. The findings suggest that mitigation strategies for AXI-based FPGA systems should consider not only configuration-memory correction, but also placement and routing isolation, frame-aware monitoring, transaction replay, and golden-reference-based detection of silent data corruption.

Future work will extend this approach to additional workloads, clock frequencies, AXI topologies, and FPGA families. Increasing RO spatial coverage would improve observability of routing perturbations beyond the selected frame regions monitored in this study. Future experiments should also evaluate mitigation schemes such as isolation-aware placement, selective scrubbing, transaction replay, and adaptive clocking to determine which techniques most effectively reduce the propagation of routing-induced disturbances into AXI-level failures.

## ACKNOWLEDGMENT

The author would like to thank the École de technologie supérieure (ÉTS), Department of Electrical Engineering, and CMC Microsystems, Kingston, ON, Canada, for providing access to advanced design tools. The author also gratefully acknowledges the assistance of the neutron-beam facility staff at TRIUMF laboratory in British Columbia, Canada, during the irradiation campaign. This work was conducted without external financial support, and the neutron irradiation experiments were funded directly by the author.

## REFERENCES

[1] M. Darvishi, Y. Audet, Y. Blaquiere, C. Thibeault, and S. Pichette, "On the Susceptibility of SRAM-Based FPGA Routing Network to Delay Changes Induced by Ionizing Radiation," *IEEE Trans. Nucl. Sci.*, vol. 66, no. 3, pp. 643–654, Mar. 2019, doi: 10.1109/TNS.2019.2898894.
[2] I. Souvatzoglou *et al.*, "Assessing the Reliability of FPGA-Based Quantized Neural Networks Under Neutron Irradiation," *IEEE Trans. Nucl. Sci.*, vol. 71, no. 12, pp. 2565–2577, Dec. 2024, doi: 10.1109/TNS.2024.3491503.

[3] M. Darvishi, Y. Audet, and Y. Blaquiere, "Delay Monitor Circuit and Delay Change Measurement Due to SEU in SRAM-Based FPGA," *IEEE Trans. Nucl. Sci.*, vol. 65, no. 5, pp. 1153–1160, May 2018, doi: 10.1109/TNS.2018.2828785.
[4] M. Darvishi, Y. Audet, Y. Blaquiere, C. Thibeault, S. Pichette, and F. Z. Tazi, "Circuit Level Modeling of Extra Combinational Delays in SRAM-Based FPGAs Due to Transient Ionizing Radiation," *IEEE Trans. Nucl. Sci.*, vol. 61, no. 6, pp. 3535–3542, Dec. 2014, doi: 10.1109/TNS.2014.2369424.
[5] C. Thibeault *et al.*, "On Extra Combinational Delays in SRAM FPGAs Due to Transient Ionizing Radiations," *IEEE Trans. Nucl. Sci.*, vol. 59, no. 6, pp. 2959–2965, Dec. 2012, doi: 10.1109/TNS.2012.2222668.
[6] J. C. Fabero *et al.*, "Single Event Upsets Under 14-MeV Neutrons in a 28-nm SRAM-Based FPGA in Static Mode," *IEEE Trans. Nucl. Sci.*, vol. 67, no. 7, pp. 1461–1469, Jul. 2020, doi: 10.1109/TNS.2020.2977874.
[7] C. De Sio, S. Azimi, and L. Sterpone, "On the Evaluation of SEU Effects on AXI Interconnect Within AP-SoCs," in *Architecture of Computing Systems – ARCS 2020*, vol. 12155, A. Brinkmann, W. Karl, S. Lankes, S. Tomforde, T. Pionteck, and C. Trinitis, Eds., in Lecture Notes in Computer Science, vol. 12155. , Cham: Springer International Publishing, 2020, pp. 215–227. doi: 10.1007/978-3-030-52794-5_16.
[8] C. De Sio, S. Azimi, and L. Sterpone, "On the analysis of radiation-induced failures in the AXI interconnect module," *Microelectronics Reliability*, vol. 114, p. 113733, Nov. 2020, doi: 10.1016/j.microrel.2020.113733.
[9] A. Perez-Celis, C. Thurlow, and M. Wirthlin, "Emulating Radiation-Induced Multicell Upset Patterns in SRAM FPGAs With Fault Injection," *IEEE Trans. Nucl. Sci.*, vol. 68, no. 8, pp. 1594–1599, Aug. 2021, doi: 10.1109/TNS.2021.3071704.
[10] D. Agiakatsikas *et al.*, "Single Event Effects Assessment of UltraScale+ MPSoC Systems Under Atmospheric Radiation," *IEEE Trans. Rel.*, vol. 73, no. 1, pp. 771–783, Mar. 2024, doi: 10.1109/TR.2023.3312548.
[11] D. S. Lee *et al.*, "Single-Event Characterization of 16 nm FinFET Xilinx UltraScale+ Devices with Heavy Ion and Neutron Irradiation," in *2018 IEEE Nuclear & Space Radiation Effects Conference (NSREC 2018)*, Waikoloa Village, HI: IEEE, Jul. 2018, pp. 1–8. doi: 10.1109/NSREC.2018.8584313.
[12] A. Stoddard, A. Gruwell, P. Zabriskie, and M. J. Wirthlin, "A Hybrid Approach to FPGA Configuration Scrubbing," *IEEE Trans. Nucl. Sci.*, vol. 64, no. 1, pp. 497–503, Jan. 2017, doi: 10.1109/TNS.2016.2636666.
[13] H. Quinn, "Challenges in Testing Complex Systems," *IEEE Trans. Nucl. Sci.*, vol. 61, no. 2, pp. 766–786, Apr. 2014, doi: 10.1109/TNS.2014.2302432.
[14] M. Preston *et al.*, "Proton- and Neutron-Induced Single-Event Upsets in FPGAs for the PANDA Experiment," *IEEE Trans. Nucl. Sci.*, vol. 67, no. 6, pp. 1093–1106, Jun. 2020, doi: 10.1109/TNS.2020.2987173.
[15] J. C. Fabero *et al.*, "Single Event Upsets Under 14-MeV Neutrons in a 28-nm SRAM-Based FPGA in Static Mode," *IEEE Trans. Nucl. Sci.*, vol. 67, no. 7, pp. 1461–1469, Jul. 2020, doi: 10.1109/TNS.2020.2977874.
[16] R. Giordano *et al.*, "Neutron-Irradiation Testing of FPGA-Embedded Hadron Fluence Sensors," *IEEE Trans. Nucl. Sci.*, vol. 70, no. 5, pp. 774–781, May 2023, doi: 10.1109/TNS.2023.3265740.
[17] L. Sterpone and M. Violante, "A New Partial Reconfiguration-Based Fault-Injection System to Evaluate SEU Effects in SRAM-Based FPGAs," *IEEE Trans. Nucl. Sci.*, vol. 54, no. 4, pp. 965–970, Aug. 2007, doi: 10.1109/TNS.2007.904080.
[18] A. Portaluri, C. De Sio, S. Azimi, and L. Sterpone, "SEU Mitigation on SRAM-based FPGAs through Domains-based Isolation Design Flow," in *2021 21th European Conference on Radiation and Its Effects on Components and Systems (RADECS)*, Vienna, Austria: IEEE, Sep. 2021, pp. 1–4. doi: 10.1109/RADECS53308.2021.9954492.
[19] M. Akhsham and Z. Navabi, "Integrating an Interconnect BIST with Crosstalk Avoidance Hardware," in *2021 IEEE 27th International Symposium on On-Line Testing and Robust System Design (IOLTS)*, Torino, Italy: IEEE, Jun. 2021, pp. 1–6. doi: 10.1109/IOLTS52814.2021.9486702.
[20] J. D. Corbett, "The xilinx isolation design flow for fault-tolerant systems," Xilinx Inc., 2012, [Online].
[21] N. Baker, E. Campbell, A. Wilson, and M. Wirthlin, "Post-Irradiation Fault Injection for Complex FPGA Designs," *IEEE Trans. Nucl. Sci.*, vol. 72, no. 8, pp. 2919–2927, Aug. 2025, doi: 10.1109/TNS.2025.3566344.
[22] M. Cannon, A. Keller, and M. Wirthlin, "Improving the Effectiveness of TMR Designs on FPGAs with SEU-Aware Incremental Placement," in *2018 IEEE 26th Annual International Symposium on Field-Programmable Custom Computing Machines (FCCM)*, Boulder, CO: IEEE, Apr. 2018, pp. 141–148. doi: 10.1109/FCCM.2018.00031.
[23] H. Quinn and M. Wirthlin, "Validation Techniques for Fault Emulation of SRAM-based FPGAs," *IEEE Trans. Nucl. Sci.*, vol. 62, no. 4, pp. 1487–1500, Aug. 2015, doi: 10.1109/TNS.2015.2456101.
[24] H. Quinn *et al.*, "Using Benchmarks for Radiation Testing of Microprocessors and FPGAs," *IEEE Trans. Nucl. Sci.*, vol. 62, no. 6, pp. 2547–2554, Dec. 2015, doi: 10.1109/TNS.2015.2498313.
[25] Xilinx Inc., "UltraScale Architecture Clocking Resources User Guide (UG572)," 2013. [Online].
[26] C. Cazzaniga, R. G. Alia, M. Kastriotou, M. Cecchetto, P. Fernandez-Martinez, and C. D. Frost, "Study of the Deposited Energy Spectra in Silicon by High-Energy Neutron and Mixed Fields," *IEEE Trans. Nucl. Sci.*, vol. 67, no. 1, pp. 175–180, Jan. 2020, doi: 10.1109/TNS.2019.2944657.
[27] C. Chen, J. Wang, X. Hu, and S. Liu, "A Hybrid RO-TDL-Based On-Chip Voltage Monitor for FPGA Applications," *IEEE Trans. VLSI Syst.*, vol. 33, no. 5, pp. 1384–1395, May 2025, doi: 10.1109/TVLSI.2024.3509439.
[28] M. Fathy, H. Nassar, M. Abd El Ghany, and J. Henkel, "Timekeepers: ML-Driven SDF Analysis for Power-Wasters Detection in FPGAs," *ACM Trans. Embed. Comput. Syst.*, vol. 24, no. 5s, pp. 1–26, Sep. 2025, doi: 10.1145/3761809.
[29] J. H. Anderson and F. N. Najm, "Low-Power Programmable FPGA Routing Circuitry," *IEEE Trans. VLSI Syst.*, vol. 17, no. 8, pp. 1048–1060, Aug. 2009, doi: 10.1109/TVLSI.2009.2017443.